\documentclass[10pt,conference]{IEEEtran}
\usepackage{cite}
\usepackage{amsmath,amssymb,amsfonts}
\usepackage{algorithmic}
\usepackage{graphicx}
\usepackage{textcomp}
\usepackage{xcolor}
\usepackage[hyphens]{url}
\usepackage{fancyhdr}
\usepackage[colorlinks=true, linkcolor=black, citecolor=black, filecolor=black, urlcolor=black]{hyperref}

\usepackage{amsmath,amssymb,amsfonts}
\usepackage{graphicx}
\usepackage{textcomp}
\usepackage{tikz}

\usepackage{booktabs}
\usepackage{xfrac}
\usepackage{enumerate}
\usepackage[shortlabels]{enumitem}
\usepackage{setspace}
\usepackage{multirow}

\usepackage{siunitx}

\usepackage[ruled,linesnumbered]{algorithm2e}
\usepackage{xcolor}
\usepackage{tikz}
\usetikzlibrary{fit,calc}
\usepackage{mathtools, nccmath}
\usepackage{algorithmic}

\usepackage{graphicx}
\usepackage{caption}
\usepackage{subcaption}

\usepackage[]{hyperref}
\usepackage{tabularx}

\SetKwInput{KwInput}{Input}
\SetKwInput{KwOutput}{Output}

\newcounter{daggerfootnote}
\newcommand*{\daggerfootnote}[1]{%
    \setcounter{daggerfootnote}{\value{footnote}}%
    \renewcommand*{\thefootnote}{\fnsymbol{footnote}}%
    \footnote[2]{#1}%
    \setcounter{footnote}{\value{daggerfootnote}}%
    \renewcommand*{\thefootnote}{\arabic{footnote}}%
    }

\newcolumntype{C}[1]{>{\centering\arraybackslash}p{#1}}  

\begin{document}

\pdfpagewidth=8.5in
\pdfpageheight=11in

\newcommand{\hpcayear}{2025}

\newcommand{\hpcasubmissionnumber}{1301}
\title{Osiris: A Systolic Approach to Accelerating \\
        Fully Homomorphic Encryption}

\def\hpcacameraready{} 
\newcommand{\hpcapubid}{0000--0000/00\$00.00}
\newcommand\hpcaauthors{
    \begin{tabular}{c @{\hspace{2cm}} c} 
    Austin Ebel & Brandon Reagen \\
    New York University & New York University \\
    Brooklyn, NY, USA & Brooklyn, NY, USA \\
    $\mathtt{abe5240@nyu.edu}$ & $\mathtt{bjr5@nyu.edu}$ \\
    \end{tabular}
    \vspace{-30px}
}
\newcommand\hpcaaffiliation{} 
\newcommand\hpcaemail{} 



\author{
  \ifdefined\hpcacameraready
    \IEEEauthorblockN{\hpcaauthors{}}
      \IEEEauthorblockA{
        \hpcaaffiliation{} \\
        \hpcaemail{}
      }
  \else
    \IEEEauthorblockN{\normalsize{HPCA \hpcayear{} Submission
      \textbf{\#\hpcasubmissionnumber{}}} \\
      \IEEEauthorblockA{
        Confidential Draft \\
        Do NOT Distribute!!
      }
    }
  \fi 
}

\fancypagestyle{camerareadyfirstpage}{%
  \fancyhead{}
  \renewcommand{\headrulewidth}{0pt}
  \fancyhead[C]{
    \ifdefined\aeopen
    \parbox[][12mm][t]{13.5cm}{}    
    \else
      \ifdefined\aereviewed
      \parbox[][12mm][t]{13.5cm}{}
      \else
      \ifdefined\aereproduced
      \parbox[][12mm][t]{13.5cm}{}
      \else
      \parbox[][0mm][t]{13.5cm}{}
    \fi 
    \fi 
    \fi 
    \ifdefined\aeopen 
      \includegraphics[width=12mm,height=12mm]{ae-badges/open-research-objects.pdf}
    \fi 
    \ifdefined\aereviewed
      \includegraphics[width=12mm,height=12mm]{ae-badges/research-objects-reviewed.pdf}
    \fi 
    \ifdefined\aereproduced
      \includegraphics[width=12mm,height=12mm]{ae-badges/results-reproduced.pdf}
    \fi
  }
  \fancyfoot[C]{}
}
\fancyhead{}
\renewcommand{\headrulewidth}{0pt}

\maketitle
\ifdefined\hpcacameraready 
  \thispagestyle{camerareadyfirstpage}
  \pagestyle{empty}
\else
  \thispagestyle{plain}
  \pagestyle{plain}
\fi

\newcommand{\hpcaheight}{0mm}
\ifdefined\eaopen
\renewcommand{\hpcaheight}{12mm}
\fi


\begin{abstract}

In this paper we show how fully homomorphic encryption (FHE) can be accelerated using a systolic architecture. 
We begin by analyzing FHE algorithms and then develop systolic or systolic-esque units for each major kernel. 
Connecting units is challenging due to the different data access and computational patterns of the kernels. 
We overcome this by proposing a new data tiling technique that we name limb interleaving. 
Limb interleaving creates a common data input/output pattern across all kernels that allows the entire architecture, named $\mathsf{\textbf{Osiris}}$, to operate in lockstep. 
$\mathsf{\textbf{Osiris}}$ is capable of processing key-switches, bootstrapping, and full neural network inferences with high utilization across a range of FHE parameters.
To achieve high performance, we propose a new giant-step centric (GSC) dataflow that efficiently maps state-of-the-art FHE matrix-vector product algorithms onto $\mathsf{\textbf{Osiris}}$ by optimizing for reuse and parallelism.
Our evaluation of $\mathsf{\textbf{Osiris}}$ shows it outperforms the prior state-of-the-art accelerator on all standard benchmarks.

\end{abstract}

\vspace{.5em}
\section{Introduction}

The ubiquity of cloud computing and online services has caused wide-scale data outsourcing, raising privacy and security concerns.
This has sparked an increased interest in techniques for privacy-preserving and confidential computing.
These techniques enable evaluating functions directly on encrypted data, ensuring it is always secured while off device.
One popular technique for cyptographic confidential computing is fully homomorphic encryption (FHE).
FHE provides secure outsourced computation using emerging encryption schemes, making it a natural fit for today's cloud model.
While promising, broad deployment of FHE remains limited due to the significant performance overhead it introduces.

Prior work has made significant progress in addressing these overheads spanning multiple domains.
Cryptographers have devised new FHE schemes and optimizations for better performance (e.g., SIMD ciphertext packing \cite{simd}, hoisting \cite{halevishoup, bossuat}, and CKKS's native fixed-point support \cite{ckks}).
Compiler research has assuaged many of FHE's programming complexities while simultaneously improving performance, including instruction scheduling \cite{porcupine, viand2023heco, coyote}, SIMD ciphertext packing and data layout \cite{ebel2023orion, helayers, gazelle, fhelipe}, and bootstrap placement \cite{dacapo, benhamouda2016optimization}.
Finally, work on hardware acceleration has proposed architectures that provide the computational power needed to overcome the slowdown.
These designs span the architectural landscape.
Early solutions targeted fixed-function designs to demonstrate the magnitudes of slowdown could be overcome~\cite{reagen2020cheetah, heax}.
Others proposed vector architectures, a natural fit as the core data structure in FHE (polynomial rings) readily maps to the ISAs~\cite{feldmann2021f1, craterlake, rpu}.
Recent efforts have included tiled/clustered architectures and FPGAs.
In tiled solutions, processing lanes and NoCs can be reformed to execute different FHE kernels~\cite{bts, ark, sharp}.
FPGA specific designs provide immediate speedup running on existing commercially available hardware~\cite{mad, agrawal2022fab, heax, ingrid_fpga_hpca, agrawalheap}.
The insights and reported speedup of prior work is substantial.
However, there is still room for improvement.
For example, fixed-function ASICs do not provide the flexibility to support a variety of FHE workloads nor algorithms;
vectors are flexible but rely on complex circuitry for generality and performance (e.g., register files and chaining);
tiled/clustered may be difficult to program and require large NoCs.

Systolic architectures can address many of the problems noted above.
The rigid, lockstep execution model of a systolic design eliminates the need for data sharing via register files (or chaining), NoC wiring, and promotes minimal control overhead.
At the same time, flexibility in data layout and staging provides the necessary generality to support a range of FHE parameters and workloads.
In this paper we develop a systolic architecture to accelerate FHE, named $\mathsf{Osiris}$.
The main intellectual insight is recognizing and distilling the complexity of FHE into a set of relatively simple computational cells and communication patterns, and the devising a dataflow that enables composing them into a complete architecture that runs in lockstep.

$\mathsf{Osiris}$ makes multiple contributions to accomplish this goal.
First, we propose and develop individual systolic 
\emph{units} for each FHE kernel,
see Section~\ref{sect:osiris-arch} for details.
We carefully consider how kernels interact and co-design units with those they connect with to optimize interfaces.
Effectively integrating units into a complete systolic architecture is non-trivial.
The key challenge is that the number theoretic transform (NTT) kernel naturally works over individual inputs (i.e., limbs). 
Conversely, the basis conversion (BConv) kernel, which we compute as a matrix multiply, demands limb tiling and interspersing data from distinct limbs to implement as a monolithic 2D systolic array. 
Na\"{\i}ve solutions would require large buffers between the units to reorganize data and introduce stalls.
Our second contribution is to solve this by developing \emph{limb interleaving}.

Limb interleaving tiles unit inputs by interspersing cross sections of distinct limbs for processing, rather than executing each limb entirely one at a time.
With limb interleaving, limbs can then flow directly through the NTT unit and into the BConv unit, eliminating large buffers and stalls.
Furthermore, as each unit runs in lockstep, the I/O rates are straightforward to balance.
With limb interleaving and the developed systolic units, we comprise a complete systolic architecture capable of accelerating entire FHE workloads, including bootstrapping and neural inference.
$\mathsf{Osiris}$ further supports recent optimizations to reduce bandwidth including
on-the-fly limb extension (OF-Limb), PRNG key generation \cite{craterlake, mad}, and twiddle factor decomposition \cite{otftwiddles}.

Efficient hardware is only part of the challenge.
$\mathsf{Osiris}$ provides high-performance kernel accelerators and efficient interfaces between them, leaving flexibility in how FHE is mapped to it.
Optimizing the FHE dataflow is complex and compounded by recent algorithmic advancements to computing matrix-vector products, specifically the baby-step giant-step (BSGS) algorithm \cite{halevishoup} and double-hoisting \cite{bossuat}. Both reduce the absolute number of computations in matrix-vector products,
and with double-hoisting, the results from computationally demanding kernels are shared and reused, saving work recomputing them.
This poses two challenges:
masking data movement latency with fewer computations and
deciding which data to cache on the chip.
Our third contribution is the development of the giant-step centric (GSC) dataflow, which addresses both challenges.
In GSC, we schedule kernels such that OF-Limb overlaps with off-chip (baby-step) key reads and size $\mathsf{Osiris}$ to balance their latencies.
GSC caches partials on chip, exhaustively reusing rotated (baby-step) inputs, see Section \ref{sect:dataflow} for details. 
The GSC dataflow combined with our systolic architecture facilitates high performance and eases implementation.

This paper makes the following contributions:
\begin{enumerate}
    \item We analyze FHE kernels, considering data patterns and kernel
        interaction, to develop a systolic unit for each.
    \item We propose limb interleaving and 
        show that it enables efficient unit connections (via consistent access patterns)
        to form a larger systolic architecture, named $\mathsf{Osiris}$, 
    \item We present the GSC dataflow to map state-of-the-art algorithms (BSGS and double hoisted rotations) to $\mathsf{Osiris}$.
    \item We rigorously evaluate $\mathsf{Osiris}$ with standard benchmarks, perform sensitivity studies to understand sources speedup, and conduct roofline analysis to demonstrate the design's balance.
    $\mathsf{Osiris}$ achieves state-of-the-art performance on all benchmarks, with a bootstrap (ResNet-20) speedup of $1.16\times$ ($1.07\times$) over SHARP~\cite{sharp} and average speedup of $6603\times$ over a CPU.
    We further show near linear performance scaling with increased resources at higher bandwidths,
    executing ResNet-20 in $23.8$ ms at $4$ TB/s.
\end{enumerate}

\section{Background}

\setlength{\tabcolsep}{4pt}
\renewcommand{\arraystretch}{1.2}
\begin{table}[t]
   \small
   \caption{CKKS parameter notation and descriptions.}
   \vspace{-.5em}
   \label{tab:params}
   \centering
   \begin{tabular}{cl}
   \toprule
   \textbf{Param.}                            & \textbf{Description} \\
   \midrule
   $\mathbf{m}$                               & Message, a vector of real or complex numbers. \\
   $[ \mathtt{P} ]$                           & Plaintext polynomial encoding the message, $\mathbf{m}$.\\
   $[\![ \mathtt{C} ]\!]$                     & Ciphertext encrypting the plaintext polynomial, $[ \mathtt{P} ]$.\\
   $N$                                        & Power-of-two polynomial ring degree. \\ 
   $n$                                        & Length (slots) of the vector message, $n \leq \sfrac{N}{2}$ (or $N$). \\
   $L$                                        & Maximum multiplicative level of $[\![ \mathtt{C} ]\!]$. \\
   $\ell$                                     & Current multiplicative level. \\
   $Q$                                        & Initial polynomial modulus.\\
   $q_i$                                      & Small moduli in RNS decomposition of $Q = \prod_{i=0}^{L} q_i$. \\
   $\alpha$                                   & Limbs in hybrid decomposition basis of $Q$. \\
   $\mathtt{dnum}$                            & Decomposition number in key-switching, $\lceil \sfrac{(\ell + 1)}{\alpha} \rceil$. \\
   $P$                                        & Auxiliary modulus used in key-switching. \\
   $p_i$                                      & Small moduli in RNS decomposition of $P = \prod_{i=0}^{\alpha-1} p_i$. \\
   $L_{\text{boot}}$                          & Number of levels consumed by bootstrapping. \\
   $L_{\text{eff}}$                           & Maximum achievable level after bootstrapping. \\
   \bottomrule
   \end{tabular}
   \label{tab:ckks}
   \vspace{-2em}
\end{table}

In this section, we introduce the relevant parameters, notation, and operations of CKKS~\cite{ckks}.
An overview of the notation is shown in Table \ref{tab:ckks}. 
When possible, we adopt terminology from prior work \cite{bts, sharp, mad}. 
CKKS supports several homomorphic operations on and between plaintext polynomials and ciphertexts, similar to prior HE schemes such as BGV \cite{brakerski2014leveled} and BFV \cite{brakerski2011fully}. 
Specifically, we denote the additions (multiplications) of two ciphertexts as $\mathsf{HAdd}$ ($\mathsf{HMult}$) and additions (multiplications) of a ciphertext and a plaintext polynomial as $\mathsf{PAdd}$ ($\mathsf{PMult}$). 
We also denote the rotation of a cleartext (ciphertext) by $\mathsf{Rot}$ ($\mathsf{HRot}$). 
To perform these operations, a vector message, $\mathbf{m}$, is first encoded into a cyclotomic polynomial $[ \mathtt{P} ] = \mathbb{Z}_Q[X] / (X^N + 1)$, with coefficients modulo $Q$ and ring degree $N$. 
A ciphertext, $[\![ \mathtt{C} ]\!]$, is a pair of polynomials that can be generated from $[ \mathtt{P} ]$ and is made secure by adding a small amount of random noise
to one of the polynomials.

Since $Q$ is often large (on the order of thousands of bits), it is most efficient to decompose a polynomial into many separate polynomials, or \textit{limbs}, each with a smaller size modulus, $q_i$, such that $Q = \prod_{i=0}^{L} q_i$, using the Chinese remainder theorem \cite{cheon2019full}.
$L$ sets the ciphertext's maximum \textit{level} and operations such as $\mathsf{HMult}$ consume levels.
When the current level, $\ell$, reaches zero, an expensive $\mathsf{Bootstrap}$ procedure is required to increase the ciphertext's level to enable further computation. 
Bootstrapping consumes a fixed number of levels, $L_\text{boot}$, and therefore a ciphertext can only reach an \textit{effective} level, $L_\text{eff} = L - L_\text{boot}$ after bootstrapping.

Both $\mathsf{HMult}$ and $\mathsf{HRot}$ operations transform a ciphertext originally decrypted by a secret key, $s$, into a ciphertext only decryptable by a new secret key, $s'$. The most efficient approach to return to $s$ and enable further computation is to multiply the ciphertext with a special \textit{switching} key, $\mathsf{swk}$, to perform a re-encryption of the ciphertext under the secret key, $s$. However, a na\"{\i}ve multiplication with the $\mathsf{swk}$ adds prohibitively large noise. 
Techniques proposed by Fan and Vercauteren \cite{fan2012somewhat} and later by Han and Ki \cite{han2020better} address this issue directly, with a common theme being that they first $\mathsf{Decompose}$ a ciphertext into many \textit{digits} and perform this $\mathsf{KeyMult}$ operation at a higher modulus. 
The act of changing a ciphertext's modulus is expensive, requiring many (inverse) number theoretic transforms (I/NTTs) and basis conversion ($\mathsf{BConv}$) operations. 
In $\mathsf{Osiris}$, we collectively refer to changing the RNS representation of a modulus from $\alpha$ limbs to $\beta$ limbs as $\mathsf{ModChange}$. 
When $\alpha < \beta$ ($\alpha > \beta \hspace{0.1em}$), we refer to the operation as $\mathsf{ModUp}$ ($\mathsf{ModDown}$). 
The $\mathsf{Rescale}$ operation is a specific case of $\mathsf{ModDown}$ when $\beta = \alpha -1$. 
Finally, we adopt the hybrid variant of key-switching \cite{han2020better}, similar to existing FHE accelerators \cite{craterlake, ark, sharp}, and refer readers to prior work \cite{neda2023ciflow, bossuat, mad} for more detail.

\setlength{\textfloatsep}{0pt}
\begin{algorithm}[t!]
    \small
    \SetKwFor{For}{for}{do}{}
    \SetInd{0.5em}{0.8em}
    \setstretch{1.15}
    \caption{$\mathsf{ModChange}_{\alpha \rightarrow \beta}([\mathtt{A}])$}
    \label{alg:modswitch}
    \DontPrintSemicolon
    \KwInput{$[ \mathtt{A} ]$, a polynomial of $\alpha$ limbs with modulus $Q$.}
    \KwOutput{A polynomial with $\beta$ limbs and modulus $Q'$.}

    $[ \mathtt{A_{coeff}} ] \leftarrow \mathsf{INTT}([ \mathtt{A}]_{q_{i}})_{\hspace{0.1em}0 \leq i < \alpha} $ \;
    $[ \mathtt{B_{coeff}} ] \leftarrow \mathsf{BConv}_{\alpha \rightarrow \beta} ([\mathtt{A_{coeff}} ])$\;

    \Return $\mathsf{NTT}([ \mathtt{B_{coeff}} ]_{q'_{i}})_{\hspace{0.1em}0 \leq i < \beta}$ 
 
\end{algorithm}

Algorithm \ref{alg:modswitch} describes this $\mathsf{ModChange}$ operation. 
Note that separate I/NTTs must be applied to each limb and that the modulus changes from $Q = \prod_{i=0}^{\alpha-1} q_i$ to $Q' = \prod_{i=0}^{\beta-1} q'_i$. We also emphasize that the core operation in $\mathsf{BConv}$ a matrix-matrix multiplication between a \textit{base table} of scaling factors and the input polynomial.
In Section \ref{sect:osiris-arch}, we show how to accelerate this matrix multiplication with a $2$D systolic array.

\section{FHE Matrix-Vector Products}
\label{sect:mv-products}

We now introduce how fast FHE matrix-vector products are implemented.
We start with the na\"{\i}ve approach and then describe two compounding optimizations: 
the \textit{baby-step giant-step} (BSGS) algorithm and $\textit{hoisting}$. 
To the best of our knowledge, $\mathsf{Osiris}$ is the first ASIC to adopt state-of-the-art hoisting techniques \cite{halevishoup, bossuat} for matrix-vector products.
Matrix-vector products can account for $70\%$ to $99\%$ of the private inference latency in FHE~\cite{ebel2023orion}. Therefore, we tailor specific mapping strategies for this algorithm on $\mathsf{Osiris}$.

\begin{figure}
    \subfloat[The diagonal method. 
                $n=6$ ciphertext rotations are required, 
                one per non-zero diagonal (including trivial rotation by $0$).
                ]{%
            \label{fig:diagonal}
            \includegraphics[clip,width=\columnwidth]{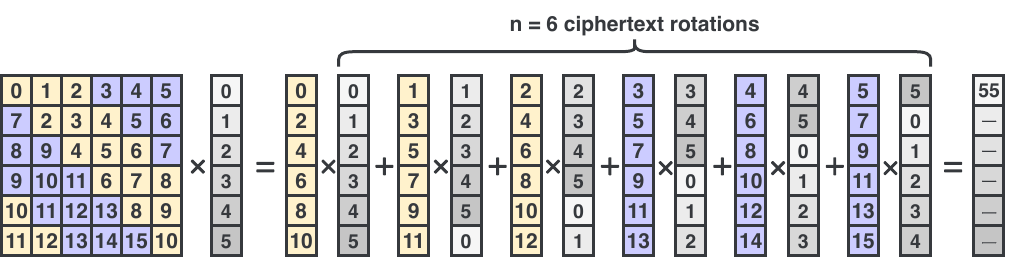}%
    }
    \vspace{5pt}
    \subfloat[Extending the diagonal method with BSGS. 
                Note that only $n_1 + n_2 = 5$ rotations are required and that $n_1 n_2 = n = 6$.
                ]{%
            \label{fig:bsgs}
            \includegraphics[clip,width=\columnwidth]{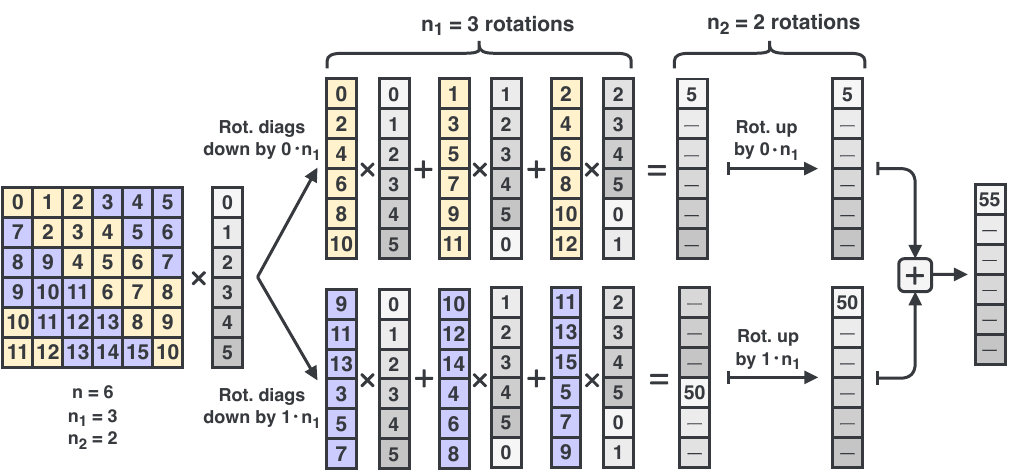}%
    }
    \caption{Visualizing how the BSGS algorithm reduces the number of ciphertext rotations in matrix-vector products.}
    \label{fig:mv-prods}
\end{figure}

\noindent\textbf{Diagonal Method:}
The simplest approach to FHE matrix-vector products is the diagonal method \cite{algsinhelib}.
Here, one first extracts the $\textit{generalized diagonals}$ of a matrix, $\mathbf{M}$. 
That is, diagonals of the form $\mathsf{diag}_k = \mathbf{M}_{[0,k]}$, $\mathbf{M}_{[1, 1+k]},$ $\ldots, \mathbf{M}_{[w-1, w-1+k]}$, where $w$ is the width of the matrix and the second index is taken modulo $w$. 
Each diagonal is multiplied with a rotated input ciphertext.
Figure \ref{fig:diagonal} shows the approach, where the number of non-zero diagonals, $n$, is 6. 
Note that since we have rotated the ciphertext $n$ times, we must perform $n$ distinct key-switch operations, each computing a $\mathsf{Decompose}$, $\mathsf{KeyMult}$, and $\mathsf{ModDown}$,
to realign partial products under the same secret key.
Thus, the runtime can be seen as $\mathcal{O}(n)$.

\noindent\textbf{Baby-Step Giant-Step:}
The baby-step giant-step (BSGS) algorithm reduces the runtime complexity of homomorphic matrix-vector products to $\mathcal{O}(\sqrt{n})$ \cite{halevishoup}. 
Therefore, as the number of non-zero diagonals increases, so does the impact of this optimization. 
We show the BSGS approach in Algorithm \ref{fig:bsgs} and modify the running matrix-vector example from Figure \ref{fig:diagonal}. 
Here, matrix-vector products are parameterized by two constants: $n_1$, the number of baby steps or input rotations, and $n_2$ the number of giant steps or partial product rotations, where $n_1 n_2=n$. 
In Figure \ref{fig:bsgs}, we set $n_1=3$ and $n_2=2$ such that $n_1 n_2 = 6$. The distinction between $n_1$ and $n_2$ lies in the magnitude of their rotations and when the rotations are performed.
We first generate the $n_1$ input rotations (baby steps) of the input ciphertext.
Each baby-step rotation is then multiplied with $n_2$ diagonals.

\begin{algorithm}[t]
    \small
    \SetKwFor{For}{for}{do}{}
    \SetInd{0.5em}{0.8em}
    \setstretch{1.25}
    \caption{$\mathsf{BSGS \hspace{3px} matrix}\times \mathsf{vector} \hspace{3px} \mathsf{algorithm}. $\cite{halevishoup}}
    \label{alg:bsgs}
    \DontPrintSemicolon
    \KwInput{$\mathsf{ct}$ encrypting vector message, $\mathbf{m}$. $\mathbf{M}_{diag}$ the diagonals of $\mathbf{M}$, an $n \times n$ matrix with $n = n_1 n_2$.}
    \KwOutput{$\mathsf{ct}' = \mathbf{M} \times \mathsf{ct.}$}

    \For{$i = 1; i < n_1; i = i + 1$}{
        $\mathsf{ct}_i \leftarrow \mathsf{HRot}_i(\mathsf{ct})$
    }
    $\mathsf{ct}' \leftarrow (0,0)$\;
    \For{$j=0$ \KwTo $n_2; j = j + 1$}{
        $\mathbf{r} \leftarrow (0,0)$\;
        \For{$i=0, i < n_1; i = i+1$}{
            $\mathbf{r} \leftarrow \mathsf{HAdd}(\mathbf{r}, \hspace{1px} 
 \mathsf{PMult}(\mathsf{ct}_i, \hspace{1px} \mathsf{Rot}_{-n_1 \cdot j}(\mathbf{M}^{n_1 \cdot j + i}_{diag})))$ 
        }
        $\mathsf{ct}' \rightarrow \mathsf{HAdd}(\mathsf{ct}', \hspace{1px} \mathsf{HRot}_{n_1 \cdot j}(\mathbf{r}))$
    }
    \Return $\mathsf{ct}'$
\end{algorithm}

Unlike in the diagonal method, we are left with $n_2$ partial products, each offset by some multiple of $n_1$. 
Therefore, $n_2-1$ non-trivial partial product rotations
(giant steps) are required to realign ciphertexts before the final reduction.
Rotations are expensive, and the key insight of BSGS is that we can choose to reuse ciphertext rotations and $\textit{instead}$ rotate matrix diagonals (Line 7, Algorithm \ref{alg:bsgs}). 
Since matrix diagonals are known a priori (weights in convolutions or constant matrices in bootstrapping), they can be trivially rotated before being encoded. This observation reduces the number of ciphertext rotations to just $n_1 + n_2$, and rotations are minimized when $n_1=n_2 = \sqrt{n}$. 

\noindent \textbf{Hoisting:} Hoisting is a cryptographic optimization that further reduces the complexity of matrix-vector products. This technique can be applied to stages of the BSGS algorithm that share a common result. For example, rotations to the \textit{same} input ciphertext share a common $\mathsf{Decompose}$ output. Therefore, we can perform this computation only once, \textit{hoisting} it out and reusing its result for all $n_1$ baby-step rotations. We further discuss the state-of-the-art double hoisting algorithm \cite{bossuat} and its mapping onto $\mathsf{Osiris}$ in Section \ref{sect:dataflow}.

\noindent\textbf{Convolutions as Matrix-Vector Products:}
In $\mathsf{Osiris}$ we convert convolutions to their equivalent matrix-vector form to benefit from hoisting in all network layers. 
This is done by adopting the methods from $\mathsf{Orion}$ \cite{ebel2023orion} and specifically employing their $\mathsf{Hybrid}$++ method.
We refer interested readers to the paper for the details of the packing strategies.

\section{Kernel Accelerator Units}
\label{sect:osiris-arch}

In this section, we introduce the hardware components, referred to as units, that comprise $\mathsf{Osiris}$. 
We begin by establishing a consistent notation for polynomials to help contrast our design with prior work. 
Then, we discuss our choice to use a pipelined FFT architecture, NTT adaptations, and its implementation in $\mathsf{Osiris}$. 
Next, we show how $\mathsf{BConv}$ can be accelerated with a monolithic 2D systolic array. 
Finally, we discuss how $\mathsf{Osiris}$ performs automorphisms and pointwise arithmetic.

\subsection{A Running Example}
\label{subsect:running_example}

\begin{figure}
    \centering
    \includegraphics[width=\linewidth]{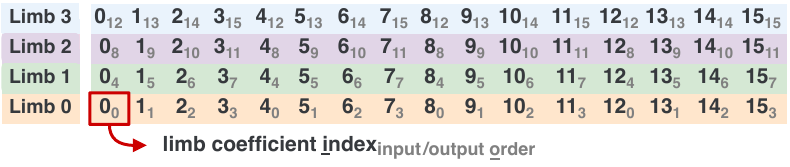}
    \caption{Matrix representation of a polynomial ($N=16$, $\ell=3$) in natural order. 
    Element $i_o$ indicates that the coefficient at index $i$ within its limb is in the $o$’th set of coefficients to be processed by a hardware unit.
    }
    \vspace{3px}
    \label{fig:ordering}
\end{figure}

Consider a polynomial with ring degree $N=16$ and multiplicative level $\ell=3$ (four limbs).
This polynomial can be represented as a $4 \times 16$ matrix, and
Figure \ref{fig:ordering} illustrates this matrix representation. 
Each element is denoted by $i_o$.
The index $i$ represents the position of the coefficient within its limb (not the coefficient itself). 
In Figure \ref{fig:ordering}, the polynomial is shown in \textit{natural} order, with indices monotonically increasing from left to right. 
The same polynomial can also be represented in \textit{bit-reversed} order, which can be distinguished by visualizing only its reordered indices. 
The subscript, $o$, indicates the order that coefficients enter or leave a hardware component. 
For example, the element highlighted in red in Figure \ref{fig:ordering} conveys that the first coefficient of the first limb is in the initial set of four values to enter or leave a unit per cycle.

\begin{figure}
    \centering
    \includegraphics[width=\linewidth]{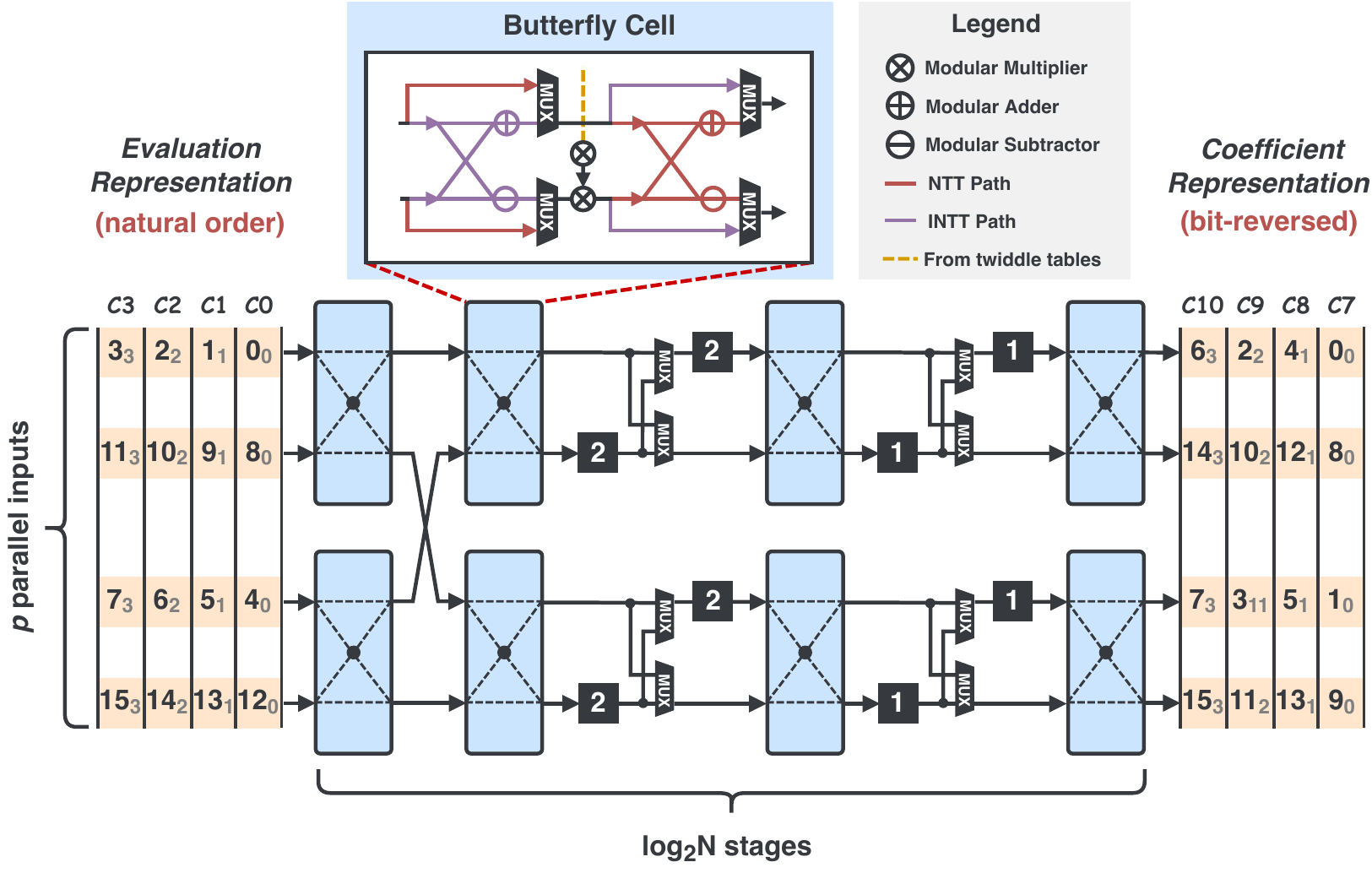}
    \caption{A $4$-parallel MDC unit (numbered boxes indicate buffer length) performing an INTT on the first limb of the example polynomial. 
    The \textit{same} input ordering shown above is compactly described by the $i_o$ notation.  
    C$0$ denotes cycle $0$.}
    \label{fig:base-mdc}
    \vspace{5px}
\end{figure}

As we introduce each of $\mathsf{Osiris}$' hardware units in the following subsections, we will refer back to this example to concretize the concepts. 
The order coefficients enter and leave functional units in $\mathsf{Osiris}$ is different from prior FHE accelerators. 
A key insight is that interleaving the computations of distinct limbs enables the lockstep execution of larger FHE kernels such as $\mathsf{ModChange}$.
Therefore, explicitly writing each coefficient in a polynomial as $i_o$ allows us to better compare this new dataflow with prior work.

\subsection{Accelerating I/NTT}

\noindent \textbf{The Multi-Delay Commutator:} The Multi-Delay Commutator (MDC) belongs to a family of pipelined FFT architectures \cite{oleary, garrido2022survey, johnston}.
In $\mathsf{Osiris}$ we employ a radix-2 pipelined MDC architecture to accelerate the (inverse) number theoretic transform (I/NTT). 
Two parameters define each MDC unit: 
$s$, the number of stages in the design, settings its depth, and 
$p$, the number of parallel inputs and outputs to each stage, defining its width.
Radix-2 pipelined architectures perform the Cooley-Tukey FFT algorithm \cite{cooleytukey}, 
which has $\log_2(N)$ stages with $\sfrac{N}{2}$ butterfly operations in each stage.
By design, each MDC stage performs \textit{all} the butterfly operations of the corresponding Cooley-Tukey stage.
Therefore, any radix-2 design will have $s=\log_2(N)$ stages. 
A $p$-parallel MDC design consists of $\sfrac{p}{2}$ butterfly \textit{cells} per stage, each computing a single two input butterfly per cycle. 
In Cooley-Tukey, the difference in coefficient indices between inputs varies from stage to stage. 
The MDC supports this by using a series of fixed buffers and multiplexers placed between each stage to \textit{commute} the order of inputs. In total, each MDC has $N-p$ buffers.
We classify the MDC architecture as systolic-\textit{esque} since all butterfly cells operate in lockstep to propagate data through the unit in a pipelined manner with minimal control logic.

\begin{figure}
    \centering
    \includegraphics[width=0.925\linewidth]{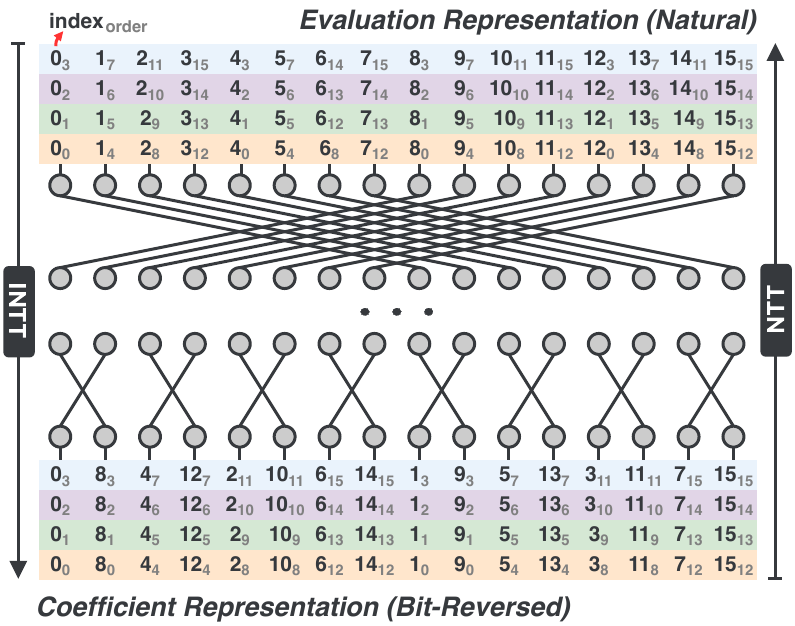}
    \caption{Visualizing the interleaved input and output order of the $4$-parallel MDC unit introduced in Figure \ref{fig:base-mdc} using $i_o$ notation introduced in Section \ref{subsect:running_example}. }
    \label{fig:mdc-order}
    \vspace{5px}
\end{figure}

\noindent \textbf{Adapting for Osiris:} $\mathsf{Osiris}$ supports a maximum ring degree of $N=2^{16}$. 
Therefore, we set $s=16$\protect\daggerfootnote{For any $N' < 2^{16}$, we skip the first $16 - \log_2(N')$ stages in the design.}.
We set the width of each MDC to be $p=512$ after carefully evaluating the bandwidth characteristics of BSGS hoisting (see Section \ref{sect:dataflow}). 
Figure \ref{fig:base-mdc} visualizes a small $4$-parallel MDC architecture performing INTT on the first limb of the example polynomial introduced in Figure \ref{fig:ordering} and highlights three key aspects of the design.
First, each butterfly cell contains \textit{two} modular multipliers, one for on-the-fly twiddle factor generation (OF-Twiddle) \cite{otftwiddles} and another for the butterfly operation itself. OF-Twiddle decomposes $N$ unique twiddle factors into two tables of size $\sqrt{N}$, enabling the generation of any twiddle factor by multiplying two entries, one from each table. This decomposition reduces the storage requirement of twiddle factors from $13$ MB to just $0.20$ MB, with groups of $p=16$ lanes sharing a common set of twiddle tables. Second, pipelined FFT architectures are bi-directional: data flowing in one direction computes the FFT, while the other direction performs the inverse FFT. 
Figure \ref{fig:base-mdc} illustrates this concept, with the highlighted butterfly cell featuring both INTT and NTT paths. 
For clarity, we omit drawing the additional control logic between stages that routes both directions \cite{4019462}. 
In Section \ref{sect:dataflow}, we show how the bi-directional nature of pipelined NTT architectures can double the throughput of the $\mathsf{DiagMult}$ kernel. 
Third, MDC architectures always accept and produce $p$ elements per cycle. 
In the following subsection, we detail how this constant flow of data simplifies rate-matching.

\noindent \textbf{Interleaving I/NTTs:} We choose the MDC design for its inherent ability \textit{interleave} the processing of independent limbs.
Recall from Algorithm \ref{alg:modswitch} that I/NTT operations occur in the $\mathsf{ModChange}$ kernel as part of the $\mathsf{INTT} \rightarrow \mathsf{BConv} \rightarrow \mathsf{NTT}$ routine. Each I/NTT acts on unique \textit{rows} (limbs) of the matrix representation.
In contrast, $\mathsf{BConv}$ is fundamentally a matrix multiplication. 
Rows of a base table are multiplied with \textit{columns} of an input polynomial. 
Each column in this polynomial contains coefficients from \textit{all} its limbs. Therefore, performing INTTs on each limb sequentially stalls the $\mathsf{ModChange}$ pipeline until all limbs have been processed.

In $\mathsf{Osiris}$, we avoid these stalls by forcing the output order of the MDC INTT to match the column-wise input order required by $\mathsf{BConv}$. 
Figure \ref{fig:mdc-order} visualizes this new \textit{interleaved} I/NTT format using the $i_o$ notation from Section \ref{subsect:running_example} and the smaller $4$-parallel MDC introduced in Figure \ref{fig:base-mdc}.
Note that while the INTT input indices are identical to those in Figure \ref{fig:ordering}, the subscripts are different. Specifically, we begin by feeding $p=4$ values from the first limb to our example MDC INTT unit. In the next cycle, rather than the next four values from the first limb, we instead feed the first four values from the second limb, beginning a second INTT. We continue this interleaving process for all limbs in the polynomial before returning to the first limb for its second set of four values. As a result, coefficients flow from the MDC INTT in the column-wise order required by $\mathsf{BConv}$, allowing the two units to operate together in lockstep.

\begin{figure}
    \centering
    \includegraphics[width=\linewidth]{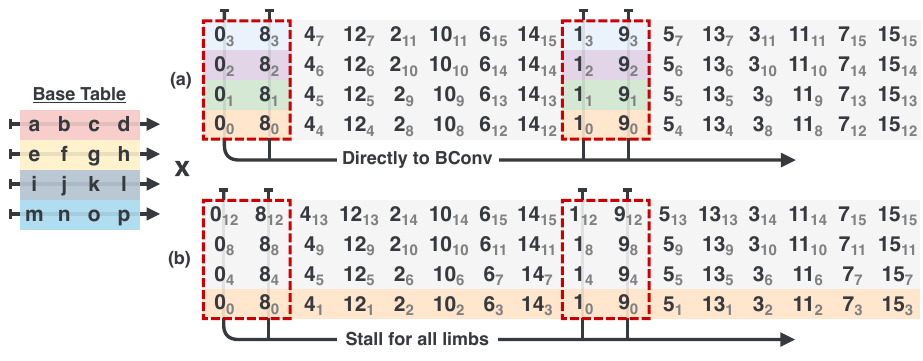}
    \caption{$\mathsf{BConv}$ input ordering in (a) $\mathsf{Osiris}$ and (b) prior FHE accelerators using the same example polynomial introduced in Section \ref{subsect:running_example}. Highlighted elements indicate coefficients available after four cycles. Note that in $\mathsf{Osiris}$, we can begin the $\mathsf{BConv}$ operation immediately since \textit{rows} of a base table are multiplied with \textit{columns} of the input polynomial.}
    \label{fig:mdc-ordering-to-bconv}
    \vspace{5px}
\end{figure}

\vspace{0.2em}
\noindent \textbf{Differences from Prior Work:} Figure \ref{fig:mdc-ordering-to-bconv} captures the differences in INTT output ($\mathsf{BConv}$ input) ordering between $\mathsf{Osiris}$ and prior FHE accelerators. Note that the $\mathsf{BConv}$ operation in $\mathsf{Osiris}$ can begin well before any single INTT completes.
To support interleaving, we simply by extend the buffers between each MDC stage by a factor equal to the number of interleaved limbs, compensating for the increased delay between butterflies of the same I/NTT. 
Therefore the $512$-parallel MDC units on $\mathsf{Osiris}$ have $13$ MB of buffers.

\begin{figure}
    \centering
    \includegraphics[width=1\linewidth]{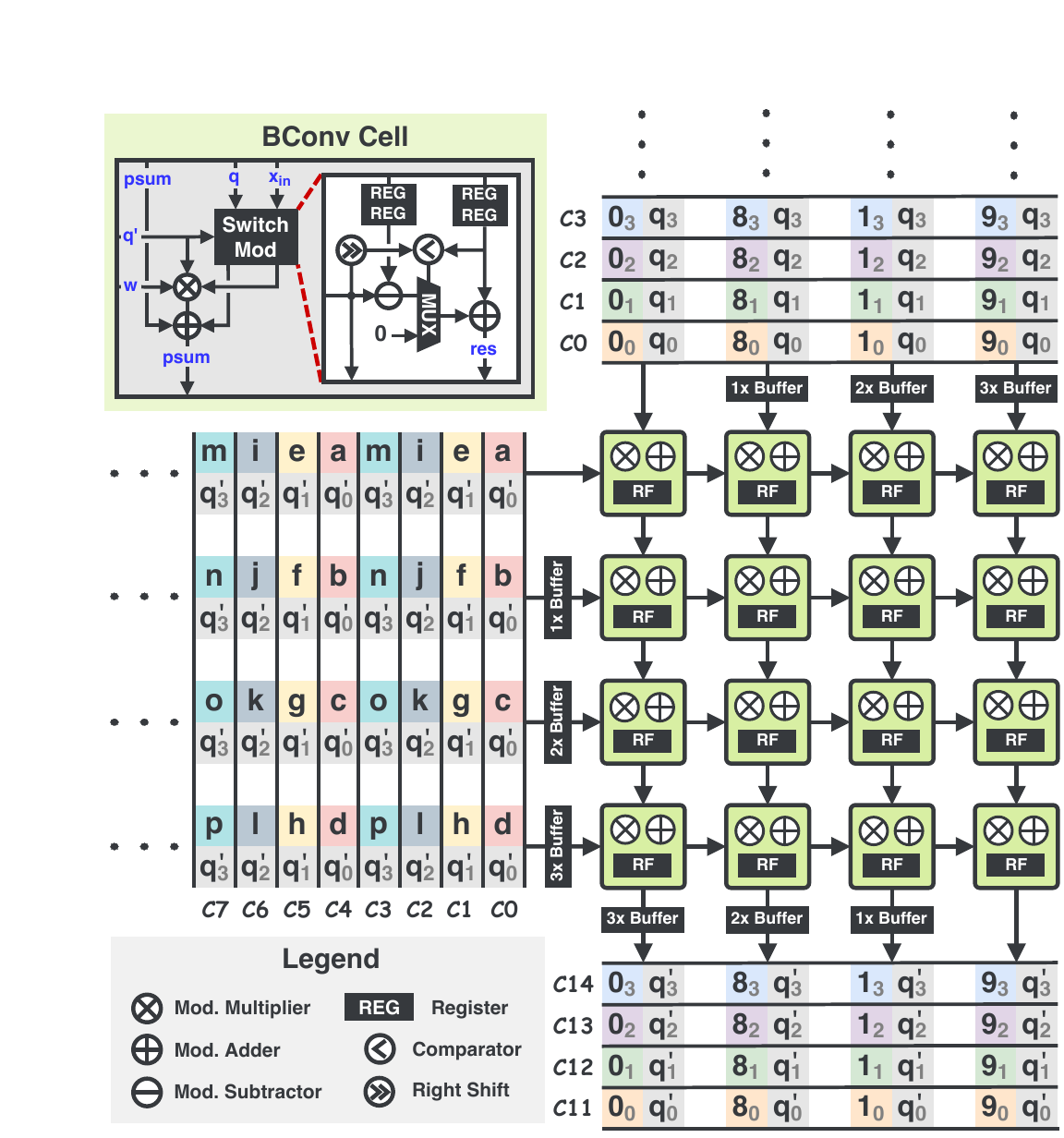}
    \caption{A $4 \times 4$ $\mathsf{BConv}$ array converting $\ell=3$ interleaved input limbs ($\hspace{-7px} \mod Q$) into $\ell=3$ interleaved output limbs ($\hspace{-7px} \mod Q'$). The first row of PEs each contain an additional modular multiplier for the preliminary limb-wise scaling in $\mathsf{BConv}$. For clarity, only the $\mathsf{psum}$ output is shown.}
    \label{fig:bconv-timing}
    \vspace{5px}
\end{figure}

\subsection{Accelerating BConv}
\noindent \textbf{2D Systolic Arrays:} The core operation of $\mathsf{BConv}$ is a large matrix-matrix multiplication, 
and it is natural to consider accelerating it with a 2D systolic array. 
Each cell performs one multiply-accumulate (MAC) operation per cycle, contributing to a partial sum that is passed down to neighboring PEs. 
Outputs stream from cells at the periphery of the array. When the array’s dimensions are smaller than the dimensions of the input matrices, blocked matrix multiplications are required.

\noindent \textbf{Adapting for Osiris:} In $\mathsf{Osiris}$, we accelerate $\mathsf{BConv}$ with a monolithic 2D input-stationary systolic array. 
The height of our array is determined by the largest height (number of limbs) of any input polynomial passed to $\mathsf{BConv}$. 
In hybrid key-switching, all $\mathsf{BConv}$ operations act on input polynomials of at most $\alpha$ limbs, either as the size of each decomposed digit in $\mathsf{ModUp}$ or as the number of limbs in the auxiliary modulus, $P$, in $\mathsf{ModDown}$ (see Table \ref{tab:params}). 
We set the maximum supported $\alpha=16$ in order to minimize the size of key-switching keys while remaining $128$-bit secure. 
Therefore, we also size the height of the systolic array to be $16$ cells tall. 
Likewise, we size the $\mathsf{BConv}$ array to be $512$ cells wide to seamlessly accept the MDC's output of $p=512$ coefficients per cycle.
As a result, the MDC naturally partitions the larger $\alpha \times N$ input polynomial into smaller blocks, each of size $\alpha \times p$, that are then fed directly to the systolic array.

Figure \ref{fig:bconv-timing} illustrates a smaller $4 \times 4$ $\mathsf{BConv}$ systolic array processing the interleaved polynomial introduced in Figure \ref{fig:mdc-order}, with $N=16$ and $\ell = 3$. 
For consistency, we use the same base table as in Figure \ref{fig:mdc-ordering-to-bconv}, converting a input polynomial from modulus $Q=\prod_{i\leq 3} {q_i}$ to a new modulus $Q'=\prod_{i\leq 3} {q'_i}$. 
We highlight four features in Figure \ref{fig:bconv-timing}. 
First, interleaved coefficients (\textit{top}) and base table constants (\textit{left}) are first skewed by buffers of differing lengths before entering the array. 
Skewed input matrices are common in 2D systolic arrays and ensure that the correct partial sums (\textit{psums}) are accumulated within the array. 
Second, input polynomial coefficients are double-buffered.
We reserve two registers within each PE for input coefficients: one for the coefficient from the current $\alpha \times p \ (4 \times 4$ in Figure \ref{fig:bconv-timing}) block being processed, and another for preloading coefficients from the next block. 
Third, all PEs perform the same \textit{switch-modulus multiply-accumulate} (SMAC) operation. For a base table constant $w$, an input coefficient $x_{in}$, and an old (new) modulus $q$ ($q'$), 
we define this operation as:
\begin{equation}
    \mathsf{psum} \leftarrow \mathsf{psum} + w \cdot \mathsf{SwitchModulus}_{q \rightarrow q'}(x_{in}) \hspace{8px} (\hspace{-10px}\mod q') \nonumber
\end{equation}

\noindent Since the $\mathsf{SwitchModulus}$ function involves both the old modulus, $q$, and the new modulus, $q'$, we stream each $q_i'$ alongside its corresponding row of base table constants. This ensures that all partial sums are accumulated under the same new modulus. 
Finally, output coefficients leave the array in an interleaved fashion and are unskewed by a set of complementary buffers. Since $\mathsf{BConv}$ is part of the larger $\mathsf{ModChange}$ pipeline, keeping coefficients interleaved ensures that they can be passed directly to subsequent MDC NTT hardware. 

\vspace{2px}

\noindent \textbf{Differences from Prior Work:} Notably, ARK \cite{ark} and SHARP \cite{sharp} employ systolic arrays to accelerate $\mathsf{BConv}$. 
Since these designs do not interleave limbs, the memory access patterns require many smaller arrays be used.
The authors allocate $1024$ $1 \times 6$ (ARK) or $2 \times 8$ (SHARP) output-stationary systolic arrays, one per lane. 
Unique base table constants are then broadcasted across the $256$ lanes in each cluster using central broadcast units (BrUs). 
In contrast, the interleaved ordering in $\mathsf{Osiris}$ enables base table constants to be passed systolically between cells in a single 2D array.

\subsection{Automorphisms}

We now describe how $\mathsf{Osiris}$ efficiently performs automorphisms. 
The automorphism applies the index permutation $\phi_r : i \rightarrow i \cdot 5^r \hspace{-4px} \mod N$ to each limb of a polynomial in the evaluation representation and in natural order. 
This permutation cyclically shifts values in the underlying vector message by $r$ slots to the left.
ARK \cite{ark} notes that with ring degree $N=2^{16}$, the automorphism can be decomposed into $2^8$ smaller permutations of $2^8$ indices. 
More generally, we can build any automorphism from permutations to subsets of coefficients if each subset, $S$, satisfies two conditions: $|S|$ is a power of two, and the indices in $S$ are evenly spaced by $\frac{N}{|S|}$. 
When $|S| = N$, we recover the original automorphism. 
This is fortuitous to $\mathsf{Osiris}$ as the MDC unit produces $p=512$ evenly spaced elements per cycle in natural order. Therefore, when an automorphism is required, the MDC's output can be directly passed to any $p$-to-$p$ permutation network, completing the operation over $\sfrac{N}{p}$ cycles. In $\mathsf{Osiris}$, we choose to perform these permutations with a $512$-to-$512$ Bene\v{s} network \cite{benes}.

\begin{figure}[t]
    \centering
    \includegraphics[width=1\linewidth]{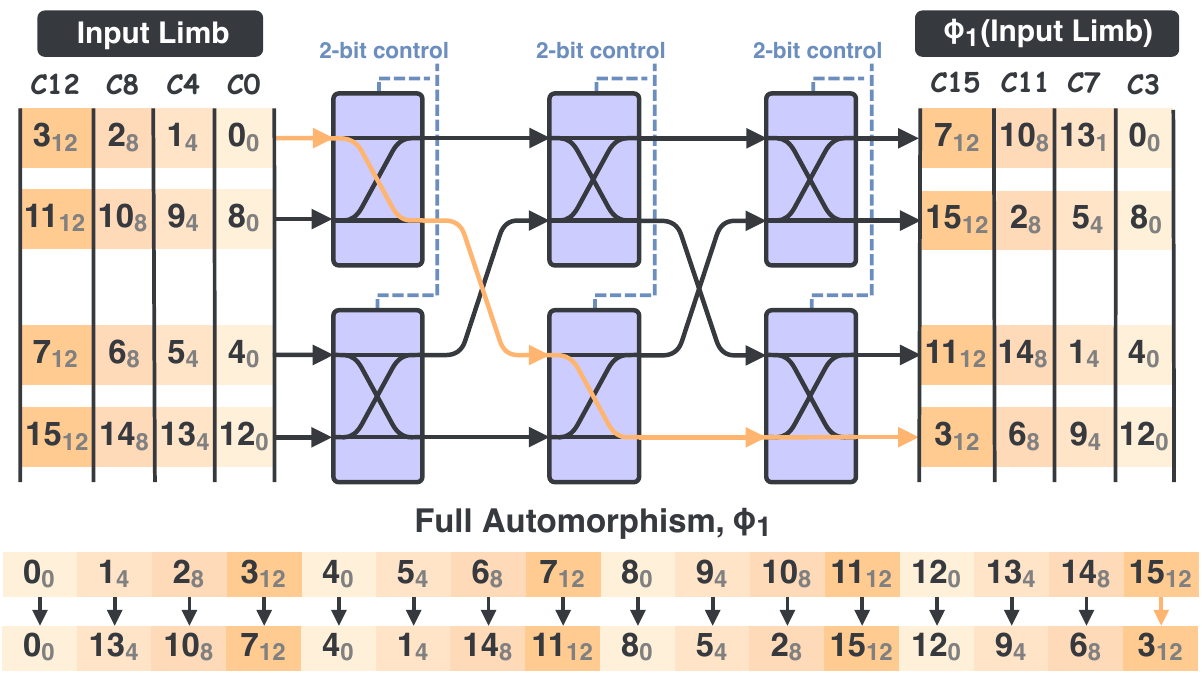}
    \caption{Using a $4$-to-$4$ Bene\v{s} network to perform the $\phi_1$ automorphism on the first limb of the interleaved, natural-ordered polynomial introduced in Figure \ref{fig:mdc-order}.}
    \label{fig:benes-network}
    \vspace{5px}
\end{figure}

Figure \ref{fig:benes-network} visualizes a smaller $4$-to-$4$ Bene\v{s} network performing $\phi_1$ on the first limb of an interleaved polynomial in natural order with $N=16$ and $\ell=3$. We highlight three features of Figure \ref{fig:benes-network}. 
First, since input limbs are interleaved, elements from the same limb enter the network once every $\ell+1=4$ cycles. 
Second, subsets of $p=4$ indices, evenly spaced by $\sfrac{N}{p} = 4$, enter the Bene\v{s} network per cycle. 
Since each subset meets the criteria above, any automorphism can be decomposed into local permutations of these subsets. 
Third, $\phi_1$ permutes each element \textit{from} index $i$ \textit{to} index $i \cdot 5^1 \hspace{-5px}\mod 16$. 
We highlight the path that maps index $3$ to index $3 \cdot 5^1 \hspace{-5px}\mod 16 = 15$, alongside the full $\phi_1$ automorphism.

\vspace{-4px}
\subsection{Hadamard Unit and Key Generation}

$\mathsf{Osiris}$ also includes a unit responsible for all point-wise multiplications and additions present in CKKS. This Hadamard unit consists of $p=512$ $1$D systolic arrays, each of height $\mathtt{dnum}$, which help to efficiently perform inner products with key-switching keys.
Each cell contains four modular multipliers and adders to support both $\mathsf{KeyMult}$ and diagonal multiplication simultaneously, which prevents stalling in our proposed dataflow.
We further employ PRNG key generation, proposed by CraterLake \cite{craterlake} and also adopted in SHARP \cite{sharp}, to generate half of all key-switching keys on-chip. 

\section{Osiris and the GSC Dataflow}
\label{sect:dataflow}

We now show how the accelerator units from Section \ref{sect:osiris-arch} are combined to build the complete $\mathsf{Osiris}$ architecture.
We begin with a discussion of on-the-fly limb generation.
Then, we present our novel \textit{giant-step centric} (GSC) dataflow for matrix-vector products and show how it can be efficiently mapped onto $\mathsf{Osiris}$.

\vspace{2px}

\noindent\textbf{On-the-fly Limb Extension:} We adopt the on-the-fly limb extension (OF-Limb) technique originally proposed for use in hardware acceleration by ARK \cite{ark} to improve arithmetic intensity. 
OF-Limb uses a single precomputed limb ($q_0$) of the plaintext diagonal (coefficient representation) to generate all other limbs.
The optimization saves significant off-chip communication at the cost of additional computation to generate limbs on-chip.
We can generate any $i$'th limb from this $q_0$-limb through the following equation:
\begin{equation*}
    [\mathtt{P}]_{q_i} = \mathsf{NTT}([\mathtt{P_{coeff}}]_{q_0}) \hspace{-4px} \mod q_i
\end{equation*}

Therefore, OF-Limb is performed with $\mathsf{Osiris}$’s MDC units. 
We use this technique to generate \textit{all} plaintext diagonals in the matrix-vector products in bootstrapping and in the linear and convolutional layers of neural networks.

\begin{figure}
    \centering
    \includegraphics[width=1\linewidth]{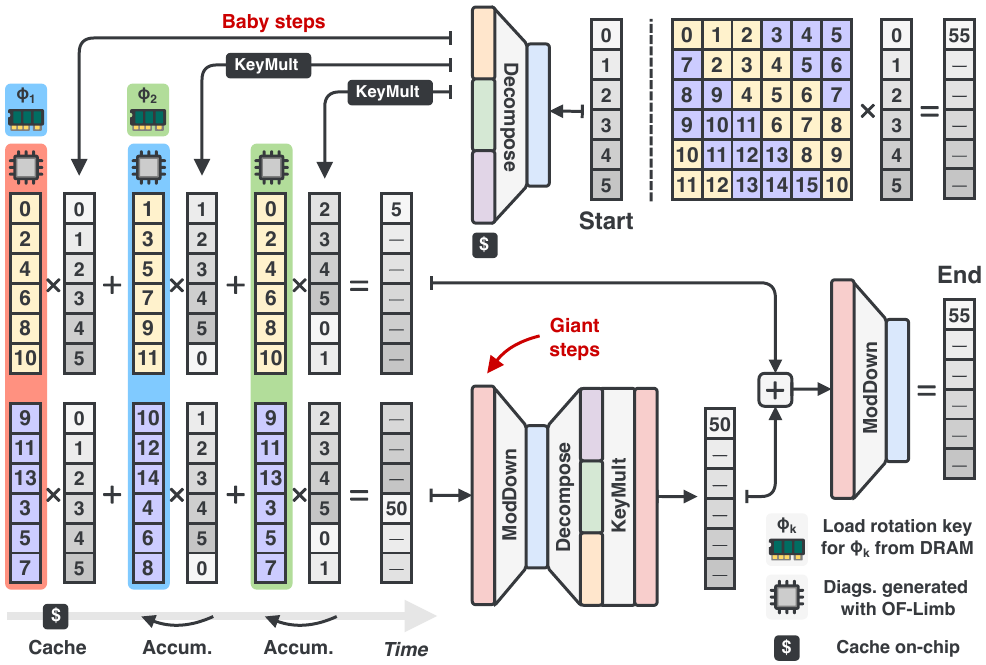}
    \caption{Visualizing how the GSC dataflow masks rotation key DRAM accesses with on-the-fly limb extension (OF-Limb) using the BSGS example introduced in Figure \ref{fig:bsgs}.}
    \label{fig:gsc-dataflow}
    \vspace{5px}
\end{figure}

\noindent \textbf{Giant-Step Centric Dataflow:} To maximize the arithmetic intensity of matrix-vector products that use the baby-step giant-step (BSGS) algorithm, we propose a novel \textit{giant-step centric} (GSC) dataflow. 
Recall from Algorithm \ref{alg:bsgs} that the BSGS algorithm reduces the number of ciphertext rotations in a matrix-vector product with $n$ non-zero diagonals from $n$ to $n_1+n_2$, where $n_1 n_2=n$. 
The goal of the GSC dataflow is both to mask the latency of loading each baby-step rotation key with diagonal generation using OF-Limb and to optimize for on-chip data reuse. 
We visualize this dataflow alongside a high-level overview of the double-hoisting algorithm in Figure \ref{fig:gsc-dataflow} using the matrix-vector product introduced in Section \ref{sect:mv-products}.  

We highlight three features of Figure \ref{fig:gsc-dataflow}. 
First, rotations to the \textit{same} input ciphertext share a common $\mathsf{Decompose}$ output. Therefore, we can perform this computation only once, \textit{hoisting} it out and reusing its result for all $n_1$ baby-step rotations. 
Similarly, Bossuat et al. \cite{bossuat} show that the $\mathsf{ModDown}$ operation commutes with addition, e.g., $\sum_i \mathsf{ModDown(ct}_i\mathsf{)} = \mathsf{ModDown}(\sum_i \mathsf{ct}_i)$. Therefore, a similar \textit{second} level of hoisting can be performed when summing all giant steps. 
We cache the $\mathsf{Decompose}$ output on-chip to leverage its significant reuse.

Second, we use OF-Limb to generate the limbs of \textit{all} plaintext matrix diagonals. Since each baby-step rotation key contributes to $n_2$ partial products, we naturally group generated diagonals into $n_1$ sets of $n_2$ diagonals to minimize off-chip memory accesses. While our MDC units are generating the current set of $n_2$ matrix diagonals, we load the next baby-step rotation key on-chip and use our Hadamard and Automorphism units to prepare the ciphertext’s next baby-step rotation. Thus, if the latency of fetching a rotation key is less than the latency of generating $n_2$ plaintext diagonals, then DRAM accesses are completely overlapped.
We deliberately size $\mathsf{Osiris}$' width with this step in mind.

Third, since we iteratively generate each set of $n_2$ partial products, we can accumulate these products on-chip, saving off-chip bandwidth and reducing the working set size. We highlight all that is cached in Figure \ref{fig:gsc-dataflow}.
When the number of giant steps that minimizes runtime
exceeds what can be stored on-chip, we find it more efficient to simply reduce the number of giant steps in the BSGS algorithm rather than send partial products off-chip. 
Since each giant step rotates a \textit{unique} ciphertext, we cannot hoist out a common $\mathsf{Decompose}$ step. 
Consequently, giant steps are more expensive than baby steps, and the optimal ratio of $n_1 / n_2$ is large, often varying between $8$ and $16$ \cite{bossuat}. 
We set the maximum number of giant steps held on-chip to be $n_2=4$ in the highest multiplicative level matrix-vector products in bootstrapping.

\begin{figure}
    \centering
    \includegraphics[width=1\linewidth]{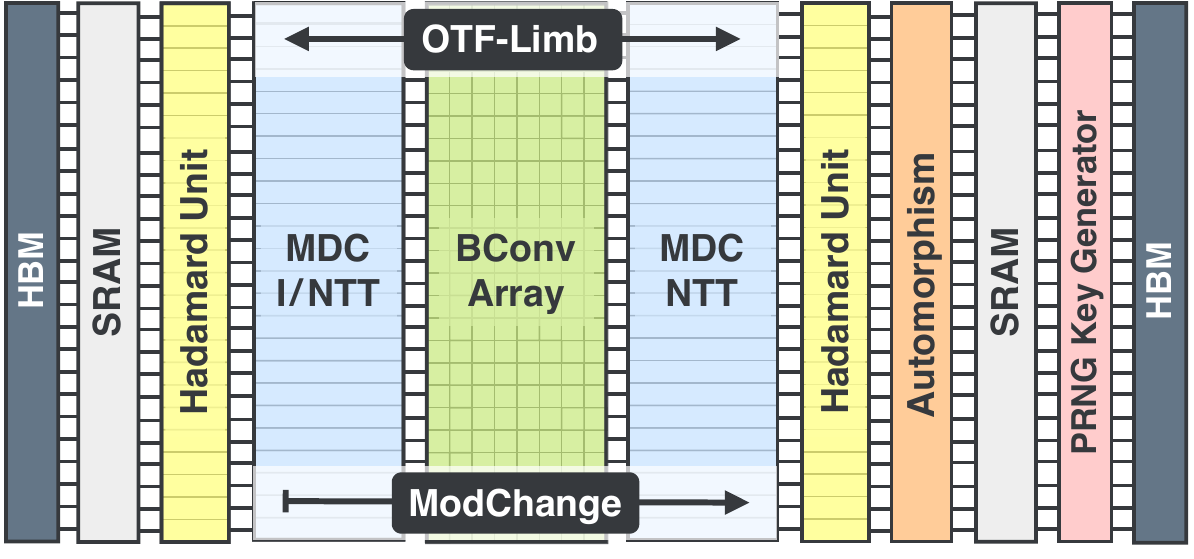}
    \caption{The complete $\mathsf{Osiris}$ architecture. Arrows indicate the flow of data through each kernel unit.}
    \label{fig:osiris-arch}
    \vspace{5px}
\end{figure}

\noindent \textbf{Architecting Osiris:}
Figure \ref{fig:osiris-arch} presents the full $\mathsf{Osiris}$ architecture. An important distinction between $\mathsf{Osiris}$ and prior FHE accelerators is that we dedicate \textit{separate} units to each step of the $\mathsf{ModChange}$ ($\mathsf{INTT}\rightarrow \mathsf{BConv} \rightarrow \mathsf{NTT}$) routine. This difference is motivated by our choice to feed limbs through $\mathsf{Osiris}$ in an \textit{interleaved} fashion. Recall from Section \ref{sect:osiris-arch} that interleaving resolves the differences between the desired input order of I/NTT and $\mathsf{BConv}$ and enables the units to operate together in lockstep. Therefore, when mapped onto $\mathsf{Osiris}$, it is useful to view the $\mathsf{ModChange}$ operation as a single, macro-pipeline with a throughput of one output limb every $\sfrac{N}{p}$ cycles.

We also note that while the BSGS algorithm reduces the number of ciphertext rotations in matrix-vector products, it does not reduce the number of plaintext diagonal multiplications. When the number of non-zero diagonals is large, a significant fraction of end-to-end latency comes from OF-Limb. To mitigate this bottleneck, we leverage the bidirectional nature of pipelined NTT architectures to reverse the direction of our $\mathsf{ModChange}$'s MDC INTT unit for the duration of OF-Limb. Since OF-Limb generates diagonals through repeated NTTs, the INTT unit would otherwise be left idle. This strategy doubles the throughput of OF-Limb, and we place Hadamard units on each side of $\mathsf{Osiris}$ to generate and accumulate partial products.

\section{Methodology}

\noindent \textbf{CPU Setup:} Alongside $\mathsf{Osiris}$, we evaluate our benchmarks on an $\mathtt{h3}$-$\mathtt{standard}$-$\mathtt{88}$ GCP instance with a Intel Xeon Platinum 8481C processor clocked at $2.7$ GHz and $352$ GB of RAM. We use Lattigo v$5.2$ to generate and evaluate our benchmarks. 

\vspace{2px}

\noindent \textbf{Incorporating Prior Work:} We support a maximum bit-width of $40$ bits and therefore adopt the same double-prime scaling techniques as SHARP \cite{sharp} to maintain high precision bootstrapping. The choice of $40$ bits over SHARP’s $36$ bits allows traditional single-prime scaling techniques to be used in the lowest level matrix-vector products in bootstrapping.  We also adopt a recent algorithmic optimization from MAD \cite{mad} to fuse the $\mathsf{ModDown}$ and $\mathsf{Rescale}$ stages within each $\mathsf{HMult}$. Its impact on performance is shown in Section \ref{sect:sensitivity}.

\vspace{2px}

\noindent \textbf{Modeling:} Given the predictability of performance in the systolic and systolic-inspired units used by $\mathsf{Osiris}$, we developed an analytical performance model. First, we tested the systolic design of each kernel with functional simulation to verify correctness. We then outputted kernel traces from Lattigo running neural inferences to appropriately model the order of operations. Kernels are mapped to corresponding units and timings for each used to estimate total runtime. 

Our goal in estimating power and area is to be commensurate with prior work.
To do this, we normalize estimates to 7nm. First, we use results from SHARP \cite{sharp} to scale the $64$-bit Montgomery multiplier units from BTS \cite{bts} to $40$ bits for an area (power) of $\SI{1236}{\micro\meter}^2$ ($\SI{0.53}{\milli\watt}$). The remaining components were synthesized using a commercial 28nm library available to us and scaled to 7nm using scaling factors of $3.6$ for area and $3.3$ for power, as assumed in prior work~\cite{28nmto16nm, 16nm,Mo_2023}.
We model power using Synopsys Design Compiler Version T-2022.03-SP2 \cite{SynopsysDC} and assume the worst case switching to conservatively model the distribution of data in cryptographic workloads.
Estimates of other logic (e.g., the Bene\v{s} network, registers, MUXs, etc.) were synthesized each using using the library and scaling down as reported above.
Our average SRAM density is $0.45$ mm$^2$/MB, which is the same as SHARP's \cite{sharp}. 
We have a total SRAM size of $184$ MB for keys, partials, decomposed polynomials, and twiddle factors;
an additional $26$ MB of space is needed for MDC buffering.
Finally, we assume the same wire overhead assumption as in ARK \cite{ark}, adding an additional $10$\% area to the total design.
We clock the chip at $1$ GHz.


\noindent \textbf{Benchmarks:} We evaluate $\mathsf{Osiris}$ on the following benchmarks: $\textbf{MLP}$ is a three-layer fully-connected network from SecureML \cite{secureML} for MNIST. $\textbf{LoLA}$ is a three-layer CNN with unencrypted weights from LoLA-CryptoNets \cite{Brutzkus2019LowLatency} also targeting MNIST and used in CraterLake~\cite{craterlake}. 
$\textbf{LeNet}$ is a four-layer CNN for MNIST (two strided convolution and two fully connected layers) adopted from $\mathsf{Orion}$ \cite{ebel2023orion} and the CHET~\cite{CHET} and EVA~\cite{EVA} FHE compilers. 
$\textbf{Bootstrap:}$ We evaluate the state-of-the-art fully-packed bootstrapping algorithm from Bossuat et al. \cite{bossuat}. $\textbf{ResNet-20:}$ We re-implement prior work of Lee et al. \cite{lee2022}, replacing convolutions with their analogous matrix-vector products, leaving all else unchanged, including the use of high-degree composite polynomial approximations to ReLU. 
By also leveraging BSGS, we reduce the number of ciphertext rotations from convolutions by $1.87\times$ over Lee et al. \cite{lee2022}. 
$\textbf{HELR-1024}$ \cite{helr1024} trains an encrypted binary classifier on a subset of MNIST images of size $14\times 14$. Each epoch performs a forward and backward pass on one mini-batch of size $1024$. Latency is reported as the average runtime per epoch. $\textbf{Sorting}$ performs bitonic sorting \cite{sorting} on $2^{14}$ elements in the range $(0,1)$. Each comparison is made by compositing twelve $7$-degree polynomials. 
We note that sparsely-packed bootstrapping (when fewer than $\sfrac{N}{2}$ slots are filled with data) is performed in sorting, HELR-1024, and ResNet-20, which is consistent with prior work.

We adopt the amortized mult time per slot metric ($\text{T}_{\text{mult} \hspace{0.1em}, \hspace{0.1em}\text{a/s.}}$)  \cite{gpuboot} and propose a similar metric that we call the \textit{amortized matrix-vector mult time per slot} defined in Equation (\ref{ammv}) for a matrix with $d$ non-zero diagonals. This metric is similar to $\text{T}_{\text{mult} \hspace{0.1em}, \hspace{0.1em}\text{a/s.}}$, however aligns more closely with order of operations found in private neural inference.

\begin{equation}
    \label{ammv}
    \text{T}_{\mathbf{M}\times v \hspace{0.1em}: \hspace{0.1em} \#d, \hspace{0.1em}\text{a/slot}} = \frac{\text{T}_{\text{boot}} + \sum_{\ell=0}^{L - L_{\text{boot}}} \text{T}_{\mathbf{M}\times v \hspace{0.1em}: \hspace{0.1em}\#d} (\ell) }{L - L_\text{boot}} \cdot \frac{2}{N}
\end{equation}
\vspace{-10px}

\setlength{\tabcolsep}{4pt}
\renewcommand{\arraystretch}{1.2}
\begin{table}[t]   
   \caption{Residual parameter sets ($\ell \leq L_\text{eff}$) used to evaluate $\mathsf{Osiris}$. Moduli ($q_i$, $p_i$) and scales ($\Delta$) are in bits. $h$ denotes the Hamming weight of each parameter set's secret key.}
   \label{tab:benchmark_params}
   \centering
   \begin{tabular}{C{0.7cm}|ccccc|ccc|c}
   \toprule
   \small
    \textbf{Set} & $N$ & $n$  & $L_\text{eff}$ & $\alpha$ & $\mathtt{dnum}$ & $q_0$ & $\Delta / q_i$ & $p_i$ & $h$\\
   \hline
    I & $2^{14}$ & $2^{14}$ & $5$ & $2$ & $3$ & $40$  & $32$ & $40$ & $192$ \\

   II & $2^{14}$ & $2^{14}$ & $7$ & $4$ & $2$ & $40$  & $32$ & $40$ & $192$ \\

   III$\textsuperscript{\textdagger}$ & $2^{16}$ & $2^{15}$ & $8$ & $5$ & $2$ & $48$ & $36$ & $40$ & $1024$ \\

   IV$\textsuperscript{\textdagger}$ & $2^{16}$ & $2^{15}$ & $9$ & $5$ & $2$ & $48$ & $36$ & $40$ & $1024$ \\

   \bottomrule
   \end{tabular}

   \vspace{1ex}
   
   \parbox{0.95\linewidth}{ 
     \footnotesize
     $\textsuperscript{\textdagger}$We apply double-prime scaling to the default bootstrapping parameters in Lattigo v5.2 \cite{lattigo}, and separately set $\alpha=14$. We further set $h=1024$ for $128$-bit security as $\log QP_{\text{max}} = 1630$; $q_0$ is split into two primes.  
   }
   \vspace{5px}
\end{table}

\section{Evaluation}
\label{sect:evaluation}

In this section, we evaluate $\mathsf{Osiris}$ on the aforementioned benchmarks and discuss the benefits afforded by systolic architectures in FHE. We begin by quantifying the implications of hoisting in hardware acceleration. Then we compare $\mathsf{Osiris}$ against prior work. Next, we perform a sensitivity study and demonstrate how $\mathsf{Osiris}$' balanced design enables near linear scaling trends with compute and bandwidth. We close with a roofline analysis \cite{williams2009roofline} to show the effects of hoisting and neural network architecture on utilization. 

\vspace{-3px}

\subsection{Benchmark Analysis}

Table \ref{tab:benchmarks} reports the latency of each benchmark using the parameter sets from Table \ref{tab:benchmark_params} on $\mathsf{Osiris}$ and with non-hoisted (NH), single-hoisted (SH), and double-hoisted (DH) BSGS matrix $\times$ vector algorithms. Table \ref{tab:benchmarks} also presents $\mathsf{Osiris}$’ improvements over the corresponding CPU implementations.

We highlight two observations in Table \ref{tab:benchmarks}. First, hoisting provides noticeable performance improvements despite lower arithmetic intensity. However, the benefits of hoisting are more pronounced at higher bandwidths. For example, double hoisting reduces the total modular multiplications in fully-packed bootstrapping on $\mathsf{Osiris}$ by $1.68 \times$ compared to BSGS without hoisting. However, at $1$ TB/s, we observe only a $1.09 \times$ improvement in latency. This discrepancy arises because we are unable to mask the latency of loading baby-step rotation keys on-chip with the computations of on-the-fly limb extension in double hoisting. As such, the design stalls for $13.9$\% of its total execution time. Conversely, BSGS without hoisting is mostly compute-limited; significantly less performance gains are achieved at higher bandwidths. 

At $2$ TB/s, the stalls in double hoisting are reduced to just $0.05$ ms, leading to a $1.47\times$ improvement over BSGS without hoisting and approaching the theoretical limit. This promising given the advent of HBM3 \cite{JESD238} and NVIDIA H100 GPUs \cite{NvidiaH100} now boasting an aggregate $8$ TB/s of bandwidth. We report latencies at $1$ TB/s to be commensurable with prior work \cite{craterlake, ark, sharp}, however we feel it is now important to consider bandwidths beyond $1$ TB/s in FHE accelerators given the pace of bandwidth scaling. We further explore scaling bandwidth and compute in Section \ref{sect:sensitivity}.

\setlength{\tabcolsep}{4pt}
\renewcommand{\arraystretch}{1.3}
\begin{table}[t!]

\centering
\caption{Benchmark latencies using non-hoisted (NH), single-hoisted (SH), and double-hoisted (DH) BSGS. }
\vspace{-0.5em}

\begin{subfigure}{\columnwidth}
\centering
\begin{minipage}{\textwidth}
\centering

\small
\label{tab:alg-results}
\centering
\begin{tabular}{C{2.3cm}|C{0.6cm}|C{0.65cm}C{0.65cm}||C{0.65cm}C{0.8cm}|C{1cm}}
\toprule
$\textbf{Benchmark}$ & \textbf{Set} & \textbf{NH} & \textbf{SH} & \textbf{DH} & \textbf{CPU} & \textbf{Improv.} \\
\hline

$\text{T}_{\text{mult} \hspace{0.1em}, \hspace{0.1em}\text{a/s.}}$ \textit{(ns)}  & III & $10.5$ & $9.89$ & $9.58$ & $68$$\mu$s & $\hspace{0.05em}\times6433$ \\ 

$\text{T}_{\mathbf{M}v \hspace{0.1em}: \hspace{0.1em} 128, \hspace{0.1em}\text{a/s.}}$ \textit{(ns)} & III  & $18.9$ & $17.3$ & $17.4$ & $112$$\mu$s & $\times \hspace{0.05em}6438$ \\ 

$\mathsf{MLP}$ \textit{($\mu$s)} & I & $124$ & $125$ & $130$ & $0.57$s & $\hspace{0.1em}\times \hspace{0.05em}4587$ \\ 

$\mathsf{LoLA}$ \textit{($\mu$s)} & I & $95.5$ & $96.7$ & $97.9$  &$0.47$s & $\hspace{0.1em}\times \hspace{0.05em}4931$ \\ 

$\mathsf{LeNet}$ \textit{($\mu$s)}  & II & $676$ & $663$ & $780$ & $2.94$s & $\hspace{0.1em}\times \hspace{0.05em}4349$ \\ 

$\mathsf{Bootstrap}$ \textit{(ms)} & III & $3.00$ & $2.82$ & $2.70$ & $17.3$s & $\hspace{0.1em}\times \hspace{0.05em}6417$\\ 

$\mathsf{HELR}$-$\mathsf{1024}$ \textit{(ms)} & III & $2.67$ & $2.61$ & $2.46$ & $15.9$s & $\hspace{0.1em}\times \hspace{0.05em}6290$\\

$\mathsf{ResNet}$-$\mathsf{20}$ \textit{(ms)} & IV & $101$ & $96.3$ & $92.4$  & $620$s & $\hspace{0.1em}\times \hspace{0.05em}7189$\\

$\mathsf{Sorting}$ \textit{(s)} & IV & $1.26$ & $1.20$ & $1.14$ & $23.0\mathsf{K}$ & $\times \hspace{0.05em}20.2\mathsf{K}$\\

\bottomrule
\end{tabular}

\end{minipage}
\vspace{-1px}
\caption{All benchmarks evaluated at $1$ TB/s, $p=512$.}
\label{fig:benchmarks}
\end{subfigure}%

\hfill

\begin{subfigure}{\columnwidth}
\centering
\begin{minipage}{\textwidth}
\centering

\small
\label{tab:alg-results}
\centering

\begin{tabular}{C{2.3cm}|C{0.6cm}|C{0.65cm}C{0.65cm}||C{0.65cm}C{0.8cm}|C{1cm}}
\toprule
$\mathsf{Bootstrap}$ \textit{(ms)} & III & $2.77$ & $2.59$ & $1.89$ & $17.3$s & $\hspace{0.1em}\times \hspace{0.05em}9153$ \\ 

$\mathsf{HELR}$-$\mathsf{1024}$ \textit{(ms)} & III & $2.49$ & $2.38$ & $2.10$ & $15.9$s & $\hspace{0.1em}\times \hspace{0.05em}7571$\\

$\mathsf{ResNet}$-$\mathsf{20}$ \textit{(ms)} & IV & $94.7$ & $89.9$ & $68.5$ & $620$s & $\hspace{0.1em}\times \hspace{0.05em}9051$\\

$\mathsf{Sorting}$ \textit{(s)} & IV & $1.20$ & $1.13$ & $0.84$ & $23.0\mathsf{K}${} & $\times \hspace{0.05em}27.4\mathsf{K}$\\
\bottomrule
\end{tabular}

\end{minipage}
\vspace{-1px}

\caption{Intensive benchmarks also evaluated at $2$ TB/s, $p=512$.}
\label{subfig:2tb-benchmarks}
\end{subfigure}%

\label{tab:benchmarks}
\end{table}

Second, we report speedup ranging from $4349\times$ to $20150\times$ at $1$ TB/s, with  benchmarks that require bootstrapping seeing the largest improvement. The higher multiplicative levels in bootstrapping entail significantly more work than operations at lower levels. $\mathsf{Osiris}$ (and systolic architectures in general) are well-suited for kernels with large, consistent streams of data.
The matrix-vector products present in bootstrapping embody this as they contain many sequential $\mathsf{BConv}$ and I/NTT operations. 
At the same time, the increased arithmetic intensity provided by on-the-fly limb extension is greatest when $\ell$ is large, as more plaintext diagonals must be generated. Therefore, while $\mathsf{Osiris}$ performs well across a spectrum of workloads, it excels in the most computationally-intense scenarios. 
The growing complexity of FHE applications is a promising trend for $\mathsf{Osiris}$.
\vspace{-5px}
\subsection{Comparisons with Prior Work}

\begin{figure}[ht]
    \centering
    \includegraphics[width=1\linewidth]{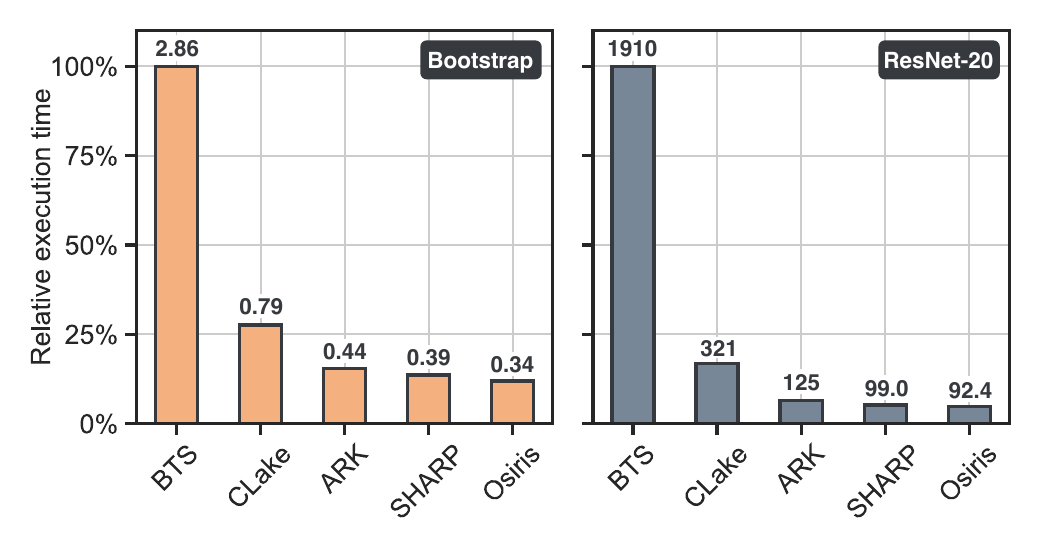}

    \vspace{-15px}
    \subfloat[$\mathsf{Bootstrap}$ / $L_\text{eff}$ (ms) \label{fig:pw-bootstrap}]{\hspace{.55\linewidth}}
    \hspace{2mm} 
    \subfloat[$\mathsf{ResNet}$-$\mathsf{20}$ (ms)\label{fig:pw-resnet}]{\hspace{.3\linewidth}}
    
    \caption{Comparison of $\mathsf{Osiris}$' $\mathsf{Bootstrap}$ (normalized by $L_\text{eff}$) and $\mathsf{ResNet}$-$\mathsf{20}$ performance versus prior work. In $\mathsf{Osiris}$, we set $L_\text{eff}=8$, similar to ARK \cite{ark} and SHARP \cite{sharp}. True latency values are shown above each bar.}
    \label{fig:pw-comparison}
    \vspace{-4px}
\end{figure}

\vspace{-1px}
\noindent\textbf{Latency:} We now compare $\mathsf{Osiris}$’ performance against the prior work of BTS \cite{bts}, CraterLake \cite{craterlake}, ARK \cite{ark}, and SHARP \cite{sharp}. Figure \ref{fig:pw-comparison} focuses on the latency of bootstrapping and $\mathsf{ResNet}$-$\mathsf{20}$ inference, since these two benchmarks are common across all designs. To account for differences in parameter sets, we normalize $\mathsf{Bootstrap}$ latency by $L_\text{eff}$.  

We highlight two features of Figure \ref{fig:pw-comparison}. First, $\mathsf{Osiris}$ achieves state-of-the-art on both benchmarks at a bandwidth of $1$ TB/s. We also expect to see better performance scaling with bandwidth than SHARP \cite{sharp}, where an alternative minimum key-switching strategy \cite{ark} (incompatible with hoisting) is used to boost arithmetic intensity. For instance, ARK \cite{ark}, using the same technique, reports only a $1.07\times$ improvement on $\mathsf{ResNet}$-$\mathsf{20}$  at $2$ TB/s, whereas $\mathsf{Osiris}$ improves by $1.35\times$.

\setlength{\tabcolsep}{4pt}
\renewcommand{\arraystretch}{1.25}
\begin{table}[ht!]
    \small
   \caption{Comparing $\mathsf{Osiris}$' hardware against prior work. Avg. power reported for $\mathsf{ResNet}$-$\mathsf{20}$ following SHARP \cite{sharp}.}
   \label{tab:comparison_sweeps}
   \centering
   \begin{tabular}{c|c|c|c|c|c|c}
   \toprule
   \multirow{2}{*}{$\textbf{\shortstack{ASIC\\Design}}$} & 
   \multirow{2}{*}{$\textbf{\shortstack{BW\\(TB/s)}}$} & 
   \multirow{2}{*}{$\textbf{\shortstack{Bit\\Width}}$} &
   \multirow{2}{*}{$\textbf{\shortstack{Mod.\\Mults.}}$} &
   \multirow{2}{*}{$\textbf{\shortstack{SRAM\\(MB)}}$} &
   \multirow{2}{*}{$\textbf{\shortstack{Area\\(mm$\boldsymbol{^2}$)}}$} & 
   \multirow{2}{*}{$\textbf{\shortstack{Power\\(W)}}$} \\
   & & & & & \\
   \hline
   BTS & $1$ & $64$ & $8192$ & $512$ & $373.6$ & $163.2$  \\
   ARK  & $1$ & $64$ & $19456$ & $588$ & $418.3$ & $281.3$   \\
   SHARP & $1$ & $36$ &  $35776$ & $198$ & $178.8$ & $94.70$ \\
   CLake$\textsuperscript{\textdagger}$ & $1$ & $28$ & $122\text{K}$ & $256$ & $222.7$  & $169.7$ \\
    \hline
   $\mathsf{Osiris}$ & $1$ & $40$ & $32256$ & $210$ & $246.9$ & $116.0$ \\
   \bottomrule
   
   \end{tabular}

   \vspace{1ex} 
   
   \parbox{0.95\linewidth}{
     \footnotesize
     $\textsuperscript{\textdagger}$Scaled from $14/12$nm to $7$nm to match $\mathsf{Osiris}$ and \cite{bts, ark, sharp}.
   }
    \label{tab:hardware-comparison}
    \vspace{5px}
\end{table}

\setlength{\tabcolsep}{3pt} 
\renewcommand{\arraystretch}{1.2}
\begin{table}[ht]
    \small
   \caption{Component-wise area (mm$^2$) and power (W) for $\mathsf{ResNet}$-$\mathsf{20}$. MDC buffers are included in I/NTT area/power. Wiring overhead is denoted as WO.}
   \label{tab:comparison_sweeps}
   \centering
   \begin{tabular}{l|c|c|c|c|c|c|c}
   \toprule
     & $\textbf{SRAM}$ &
     $\textbf{HBM}$ & 
     $\textbf{I/NTT}$ & $\textbf{BConv}$ & $\textbf{Hada.}$ & $\textbf{Auto.}$ & $\textbf{WO.}$  \\
   \hline
    Area & $83.0$ & $29.6$ & $75.8$ & $18.4$ & $14.1$ & $0.56$ & $22.5$ \\
    Power & $19.4$ & $31.8$ & $35.1$ & $14.0$ & $4.34$ & $0.80$ & $10.6$ \\
    
   \bottomrule
   
   \end{tabular}
    \label{tab:component-wise}
    \vspace{-10px}
\end{table}

Second, we normalize latencies relative to BTS \cite{bts}, which, to the best of our knowledge, is the only compared design that does not adopt the BSGS algorithm. To fairly compare with BTS, we also implement a bootstrapping variant that excludes both BSGS and hoisting. At $1$ TB/s, this variant achieves the highest multiplier utilization of $76.4$\% and a $\mathsf{Bootstrap}$ latency of $5.64$ ms, a $4.05\times$ improvement over BTS when normalized by $L_\text{eff}$. 
This demonstrates $\mathsf{Osiris}$'s flexibility in supporting the wide variety of workloads without hoisting. Finally, by adopting the larger $40$-bit moduli, rather than SHARP’s $36$-bit approach, we improve $\mathsf{ResNet}$-$\mathsf{20}$ performance by $1.2$ ms.

\noindent\textbf{Power/area:} Table \ref{tab:hardware-comparison} compares the hardware in $\mathsf{Osiris}$ against prior work. Similar to SHARP \cite{sharp}, we use optimistic $14$nm to $7$nm area ($2.8\times$) and power ($2.2\times$) scaling factors from \cite{14nm-to-7nm} to fairly compare against CraterLake \cite{craterlake}. Notably, since modular multiplier area and power scales superlinearly with bit width \cite{sharp}, CraterLake and SHARP can achieve a higher total compute capability for a similar area and power budget. A further component-wise area and power breakdown for $\mathsf{ResNet}$-$\mathsf{20}$ is shown in Table \ref{tab:component-wise}. Since we allocate $\textit{two}$ MDC I/NTT units in $\mathsf{Osiris}$, the contribution of these units to total area is $2.63\times$ higher than SHARP \cite{sharp}. In contrast, our $\mathsf{BConv}$ hardware consumes just $7.0$\% ($11.3$\%) of area (power).

\vspace{-4px}
\subsection{Sensitivity Study}
\label{sect:sensitivity}

\begin{figure}[ht]
    \centering
    \includegraphics[width=1\linewidth]{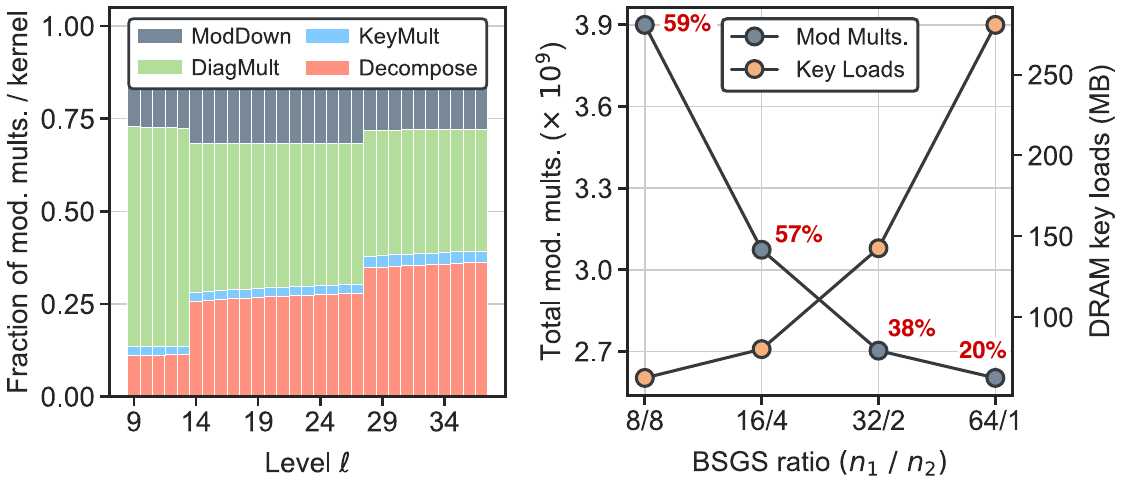}

    \vspace{-15px}
    \subfloat[\label{fig:mv-a}$L_\text{eff} \leq \ell \leq L$]{\hspace{.4\linewidth}}
    \hspace{4mm} 
    \subfloat[\label{fig:mv-b}$\ell=12$]{\hspace{.45\linewidth}}

    \caption{Exploring the impact of multiplicative level, $\ell$, and BSGS ratio on the performance of a double-hoisted matrix-vector product with $64$ non-zero diagonals. $\mathsf{DiagMult}$ includes OF-Limb. Multiplier utilization (\%) is shown in red in (b).}
    \label{fig:mv-exploration}
    \vspace{5px}
\end{figure}

\noindent\textbf{BSGS Exploration:} In Figure \ref{fig:mv-exploration}, we highlight the trade-offs that occur when evaluating matrix-vector products using the double-hoisting BSGS algorithm \cite{bossuat} on $\mathsf{Osiris}$. 

In Figure \ref{fig:mv-a}, we split the algorithm into its core kernels and evaluate a matrix-vector product with $64$ non-zero diagonals (similar to the matrices found in bootstrapping) for levels $L_\text{eff} \leq \ell \leq L$ using Set IV in Table \ref{tab:benchmark_params}. In Figure \ref{fig:mv-b}, we focus on just a single matrix-vector product at $\ell=12$, varying the BSGS ratio ($n_1 / n_2$) from $1$ to $64$ and plotting the number of modular multiplications alongside rotation key loads from DRAM. Recall from Section \ref{sect:dataflow} that a goal of the GSC dataflow is to mask the latency of loading each baby-step rotation key by generating $n_2$ matrix diagonals with OF-Limb. When the number of giant steps, $n_2$, is sufficiently large, enough diagonals can be generated to ensure that all loads are masked. For instance, in Figure \ref{fig:mv-b}, a BSGS ratio of $1$ avoids stalls entirely and maintains high multiplier utilization at $59$\%. However, when moving beyond $n_2=4$ to $n_2=1$, the design stalls for $35\%$ of its execution time, reducing utilization to just $20$\%. This increases our latency on $\mathsf{Osiris}$ by $1.96\times$ despite $1.50\times$ fewer modular multiplications. Thus, while key multiplication contributes only marginally to modular operations (Figure \ref{fig:mv-a}), it is often the deciding factor for the optimal BSGS ratio on $\mathsf{Osiris}$.

\begin{figure}
    \centering
    \includegraphics[width=1\linewidth]{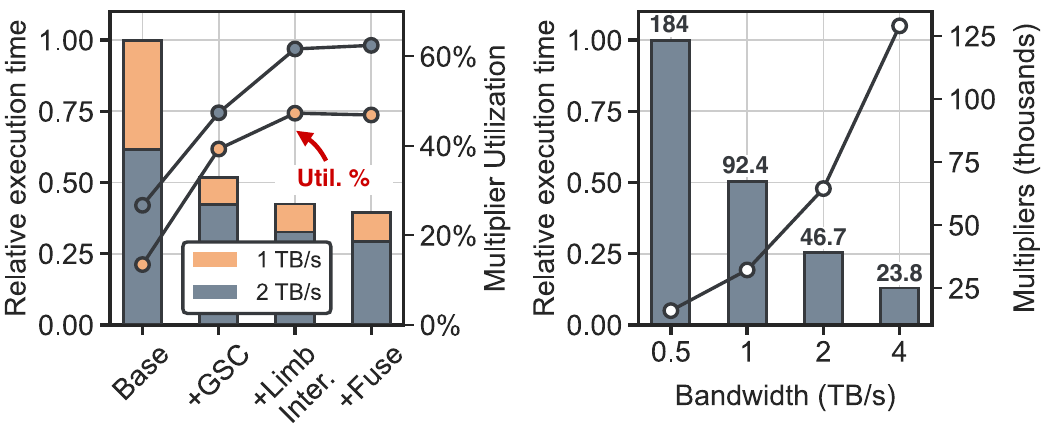}

    \vspace{-18px}
    \subfloat[Algorithm speed-up.\label{fig:resnet-alg}]{\hspace{.45\linewidth}}
    \hspace{0mm} 
    \subfloat[Resource scaling.\label{fig:resnet-bw}]{\hspace{.5\linewidth}}
    \caption{Quantifying the impact of algorithmic optimizations and bandwidth scaling on $\mathsf{Osiris}$' $\mathsf{ResNet}$-$\mathsf{20}$ latency.} 
    \label{fig:sensitivity}
    \vspace{2px}
\end{figure}

\noindent\textbf{Dataflow Optimizations:} In Figure \ref{fig:sensitivity}, we isolate the impacts of accelerator-specific algorithmic optimizations and the effects of scaling compute and bandwidth on $\mathsf{ResNet}$-$\mathsf{20}$ performance. In Figure \ref{fig:resnet-alg}, we begin with a baseline implementation of $\mathsf{Osiris}$ at $1$ TB/s with neither the GSC dataflow nor limb interleaving and build to a design that incorporates both alongside $\mathsf{ModDown}$ fusing.
By adopting the GSC dataflow for $\mathsf{ResNet}$-$\mathsf{20}$, we reduce stalls from loading rotation keys by $113$ ms ($56$ ms). Limb interleaving further boosts PE utilization from $8$\% ($14\%$) by preventing additional stalls between our I/NTT and BConv hardware in each $\mathsf{ModChange}$ operation. $\mathsf{ModDown}$ fusing further reduces $\mathsf{ResNet}$-$\mathsf{20}$ runtime by $7$\% ($8$\%). 

We also note that $\mathsf{ResNet}$-$\mathsf{20}$ latency roughly halves as we double both memory bandwidth and compute. The near linear scaling trends seen here are deliberate and follow naturally from scaling $\mathsf{Osiris}$'s array width. This highlights how the simple, lockstep execution a systolic architecture can target a variety of applications and environments with ease. 

\begin{figure}
\centering
\subfloat[Intensive benchmarks that require bootstrapping.]{%
    \label{fig:roofline1}
  \includegraphics[clip,width=0.945\columnwidth]{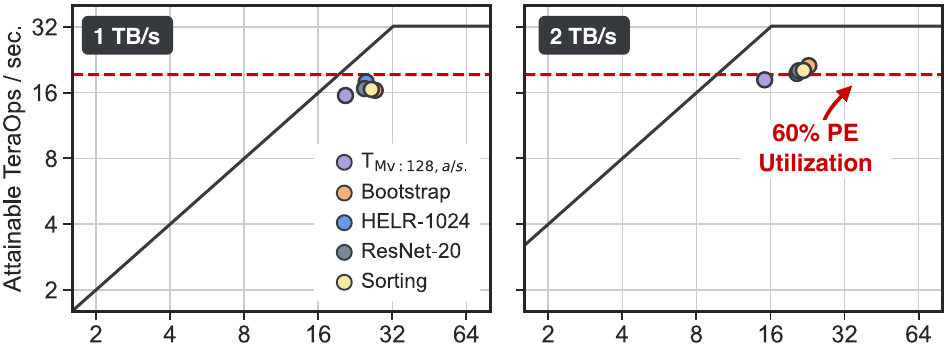}%
}

\vspace{2px}

\subfloat[In $\mathsf{ResNet}$-$\mathsf{20}$ and across matrix $\times$ vector algorithms.]{%
    \label{fig:roofline2}
  \includegraphics[clip,width=0.945\columnwidth]{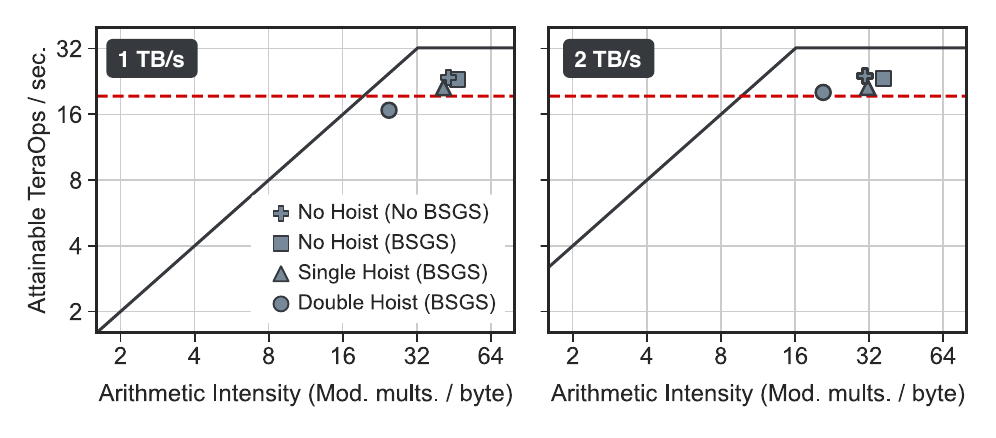}%
}

\caption{Roofline analysis of $\mathsf{Osiris}$' neural network benchmarks for varying bandwidths and matrix-vector algorithms.}
\vspace{2px}

\label{fig:roofline}
\end{figure}

\noindent\textbf{Roofline Plots:} In Figure \ref{fig:roofline}, we use roofline plots to better understand the impact of hoisting and neural network architecture on $\mathsf{Osiris}$’ performance. Roofline plots offer an intuitive way to identify sources of performance bottlenecks and inform better algorithm and hardware design. This attribute is particularly valuable in the context of FHE, where upwards of $100$ GB of rotation keys may be streamed on-chip per $\mathsf{ResNet}$-$\mathsf{20}$ inference. We define arithmetic intensity here to be the number of modular multiplications per byte of DRAM traffic; this metric has also been adopted by MAD \cite{mad}. In Figure \ref{fig:roofline1}, we visualize the characteristics of our seven benchmarks at both $1$ TB/s and $2$ TB/s. In Figure \ref{fig:roofline2}, we focus specifically on $\mathsf{ResNet}$-$\mathsf{20}$ and sweep its implementation using the matrix-vector algorithms in Section \ref{sect:dataflow}. 

We highlight three features of Figure \ref{fig:roofline}. First, $\mathsf{Osiris}$’ multiplier utilization ranges from $46$\% to $74$\% and increases with bandwidth. The extent of this increase is correlated with the amount of hoisting performed. Second, arithmetic intensity generally \textit{decreases} with increasing bandwidth. This effect occurs because we independently optimize the ratio of baby-steps to giant-steps, $n_1 / n_2$, at each bandwidth. At $2$ TB/s, the optimal ratios are often \textit{higher}, hoisting out more expensive operations and approaching the results derived in \cite{bossuat}. Third, our hardware is well-balanced and algorithms are situated near the roofline’s ridge point. This outcome is due to the deliberate sizing of $\mathsf{Osiris}$’ width and speaks to the ease at which we can reason about higher-level concepts with systolic architectures. This balanced design enables the promising scaling trends seen in Figure \ref{fig:resnet-bw}.

\vspace{-5px}
\section{Related Work}
\noindent$\textbf{CPU/GPU:}$ Multiple FHE software libraries are now available \cite{lattigo, openfhe, sealcrypto, heaan, algsinhelib}, providing high-performance CPU implementations.
The high computational and memory bandwidth of GPUs provide a promising alternative to CPUs for running FHE.
A collection of GPU implementations now exist.
One of the first to port FHE to GPU was cuHE~\cite{cuHE}, paving the way for future work such as $\mathtt{100x}$ \cite{gpuboot} and Cheddar \cite{cheddar}. 
Others have fine-tuned specific kernels (e.g., NTT) for GPUs achieving multiple orders of magnitude improvement~\cite{ntt_gpu}.

\noindent$\textbf{FPGAs:}$ Prior\cite{heax, agrawal2022fab, poseidon} employ FPGAs as lower power, more custom alternative to GPUs, and further accelerates FHE.
Notably, FAB \cite{agrawal2022fab} is the first to support CKKS bootstrapping on FPGAs and also employs the double-hoisting optimizations~\cite{bossuat}. However, while FPGA-based approaches are significantly more flexible than ASIC designs, they still lack the available on-chip memory to scale compute without bandwidth concerns. The problems above motivate the need for ASIC acceleration. HEAX proposes an architecture for ciphertext-ciphertext multiplication and reports two orders of magnitude of speedup~\cite{heax}.

\noindent$\textbf{ASIC Acceleration:}$ F1 \cite{feldmann2021f1} and Cheetah \cite{reagen2020cheetah} were among the first ASIC designs to target homomorphic encryption. Since then, many works such as CraterLake \cite{craterlake}, BTS \cite{bts}, ARK \cite{ark}, and SHARP \cite{sharp} have targeted \textit{fully} homomorphic encryption, explicitly supporting bootstrapping. CraterLake \cite{craterlake} was the first to realize sub-second ResNet-20 latencies, and optimizations such as minimum key-switching and on-the-fly limb-extension from ARK and SHARP have further reduced latencies by $3\times$. Moreover, the use of 36-bit moduli in SHARP \cite{sharp}, coupled with PRNG key generation from CraterLake \cite{craterlake}, have pushed FHE into the realm of practicality, tackling critical issues such as prohibitively large on-chip memories and low arithmetic intensity.

\section{Conclusion}

In this paper, we proposed $\mathsf{Osiris}$, a scalable systolic approach to accelerating fully homomorphic encryption. 
The benefits of systolic architectures are many: simplified hardware, intuitive dataflows, and clear compute-memory trade-offs. To complement $\mathsf{Osiris}$, we also proposed limb interleaving, a new data tiling approach that enables the entire architecture to operate in lockstep. When coupled with our giant-step centric dataflow, $\mathsf{Osiris}$ achieves state-of-the-art performance across all standard benchmarks at $1$ TB/s of bandwidth, with near linear performance improvements as bandwidth and compute are scaled.

\section*{Acknowledgement}

This work was supported in part by Graduate Assistance in Areas of National Need (GAANN). 
The research was developed with funding from the NSF CAREER award \#2340137 and from DARPA, under the Data Protection in Virtual Environments (DPRIVE) program, contract HR0011-21-9-0003. Reagen and Ebel received generous support from the NY State Center for Advanced Technology in Telecommunications (CATT) and a gift award from Google. We also thank Alhad Daftardar for the help in synthesizing our Bene\v{s} network. The views, opinions, and/or findings expressed are those of the authors and do not necessarily reflect the views of sponsors.


\bibliographystyle{IEEEtranS}
\bibliography{references}

\end{document}